\documentclass[useAMS,usenatbib]{mn2e}
\usepackage{epsf,times}

\newcommand{\bm}[1]{\mbox{\boldmath$#1$}}
\newcommand{\cnfw}{c_{\rm nfw}}
\newcommand{\nunfw}{\nu_{\rm nfw}}
\newcommand{\nulens}{\nu_{\rm lens}}
\newcommand{\nunoisy}{\nu_{\rm noisy}}
\newcommand{\nuth}{\nu_{\rm th}}
\newcommand{\numin}{\nu_{\rm min}}
\newcommand{\rvir}{r_{\rm vir}}

\title[Searching for massive clusters in weak lensing surveys]
{Searching for massive clusters in weak lensing surveys}

\author[T.~Hamana, M.~Takada \& N.~Yoshida]
{Takashi Hamana$^{1,2}$, Masahiro Takada$^3$ and Naoki Yoshida$^{2,4}$\\
$^1$ Institut d'Astrophysique de Paris, 98bis Boulevard Arago, 
F 75014 Paris, France\\ 
$^2$ National Astronomical Observatory of Japan, Mitaka, Tokyo 181-8588, 
Japan\\
$^3$ Department of Physics and Astronomy, University of Pennsylvania, 
209 S. 33rd Street, Philadelphia, PA 19104, USA\\
$^4$ Harvard-Smithsonian Center for Astrophysics, 60 Garden Street, 
Cambridge, MA 02138, USA}

\date{Accepted 30 January 2004; Received 30 January 2004; 
in original form 21 October 2003}
\pagerange{\pageref{firstpage}--\pageref{lastpage}} 
\pubyear{2003} 
\volume{000}

\begin{document}

\label{firstpage}
\maketitle

\begin{abstract}
We explore the ability of weak lensing surveys to locate massive 
clusters. We use both analytic models of dark matter halos 
and mock weak lensing surveys generated from a large cosmological
$N$-body simulation. The analytic models describe average properties 
of weak lensing halos and predict the number counts,
enabling us to compute an effective survey selection function.
We argue that the detectability of massive halos depends not
only on the halo mass but also strongly on redshift at which the 
halo is located. 
We test the model prediction for the peak number counts in weak 
lensing mass maps against the mock numerical data, 
and find that the noise due to intrinsic galaxy ellipticities 
causes a systematic effect which increases the peak counts. 
We develop a correction scheme for the systematic effect in an 
empirical manner, and show that, after the 
correction, the model prediction agrees well with the mock data.
The mock data is also used to examine the completeness and 
efficiency of the weak lensing halo search with fully taking into 
account the noise and the projection effect by large-scale 
structures. We show that the detection threshold of 
${\rm S/N}=4\sim 5$ gives an optimal balance between completeness
and efficiency.
Our results suggest that, for a weak lensing survey with a galaxy 
number density of $n_g=30$ arcmin$^{-2}$ with a mean redshift 
$z=1$,
the mean number of halos which are expected to cause lensing signals 
above ${\rm S/N}=4$ is $N_{\rm halo}({\rm S/N}>4)=37$ per 10 deg$^2$, 
whereas 23 of the halos are actually detected with ${\rm S/N}>4$,
giving the effective completeness as good as 63\%. 
On the other hand, the mean number of peaks in the same area 
is $N_{\rm peak}=62$ for a detection threshold ${\rm S/N}=4$.
Among the 62 peaks, 23 are due to halos with the 
expected peak height ${\rm S/N}>4$, 13 are due to halos with 
$3<{\rm S/N}<4$ and the remaining 26 peaks are either the 
false peaks due to the noise or halos with a lower expected peak height.
Therefore the contamination rate is 42\% (this could be an overestimation).
Weak lensing surveys thus provide a reasonably efficient way to 
search for massive clusters.
\end{abstract}

\begin{keywords}
cosmology: theory --- gravitational lensing --- galaxies: halos 
--- dark matter --- large-scale structure of universe
\end{keywords}

\section{Introduction}
\label{intro}

Recently, it has become feasible to locate massive clusters  
directly as density enhancements using weak gravitational lensing. 
Miyazaki et al.~(2002) indeed discovered many clusters in a weak lensing 
halo survey over a 2.1 square degree field, proving the ability of 
weak lensing to identify massive clusters (see also Wittman et al. 2001).
Unlike conventional optical or X-ray selected cluster catalogs, 
weak lensing cluster catalogs are free from a bias toward the luminous
objects because the cluster finding is not based on the flux enhancement
but on the {\it projected mass} enhancement.
This is a great advantage of gravitational lensing. 
However, lensing has its own disadvantages as every other cluster survey 
technique (optical, X-ray and Sunyaev-Zel'dovich survey) does;
for example, the projection effect 
from unrelated structures in the same line-of-sight 
(Reblinsky \& Bartelmann 1999; Metzler, White \& Loken 2001). 
Since clusters of galaxies are one of most important cosmological probes, 
it is of fundamental importance to construct large unbiased samples.
Given advantages and disadvantages in each survey technique, it is 
desirable to combine, in a complementary manner, several techniques 
to construct an unbiased catalog.
Such catalogs provide a valuable data set to investigate,
for example, the physical state of the intra-cluster medium.
To do this, it is important to correctly understand the 
selection function and completeness of each survey technique.

It has been often argued that weak lensing provides a truly mass-selected 
sample of clusters. This is not strictly true; weak lensing measures 
the two-dimensional tidal field which is a line-of-sight projection 
of the three-dimensional tidal field weighted by distance ratio 
$D_l D_{ls}/D_s$ (see \S 2 for definitions and details; see Mellier 1999; 
Bartelmann \& Schneider 2001 for reviews of weak lensing).
In simpler terms, the amplitude of the lensing signal from a cluster is not 
solely determined by its mass, but depends also on the redshift and the 
shape of the gravitational potential due to the cluster. 
In addition, both foreground and background large-scale structures 
contribute to the lensing signal, and this projection effect usually 
adds a noise to the lensing signal from a halo. Further, since the 
weak lensing signal is measured from tiny coherent distortions in 
galaxy shapes, intrinsic ellipticities of galaxies introduce an 
irremovable noise. Thus, the detectability of weak lensing halo 
depends on all these factors.

The primary purpose of the present paper is to explore the ability of weak 
lensing surveys to locate massive clusters.
We especially address the following three points; 
(1) to examine the selection function of weak lensing cluster surveys, 
(2) to develop a theoretical model that describes the weak lensing 
halo counts, and 
(3) to examine how the detailed structure of clusters, projection effect 
and noise affect
weak lensing cluster surveys under a typical observational condition from
a ground based $4-10$m class telescope.
The last point is especially important to understand a bias that 
weak lensing halo catalogs may have.
To address these things,
we use both simple analytic descriptions of dark matter halos
and mock numerical weak lensing survey data.
The former offers a useful way to compute expected lensing properties 
of massive halos. We use them to compute the selection function.
We also develop a theoretical model for the number counts of the 
weak lensing halos based on the analytic descriptions.
On the other hand, the latter allows us to examine {\it all}
the factors listed above in a direct manner.
Mock numerical catalogs are generated using weak lensing 
ray-tracing simulations. We extensively use the mock data and 
the halo catalogs directly produced from the simulation outputs 
to examine 
{\it completeness} and {\it efficiency} of the weak lensing halo
survey. The mock data is also used to test the model prediction 
of the weak lensing halo counts.

White, Van Waerbeke, \& Mackey (2002) and Padmanabhan, Seljak \& Pen (2003) 
addressed the ability of weak lensing surveys focusing on the projection 
effect. In those studies, the completeness of weak lensing surveys is 
defined as the fraction of detected halos with mass above a certain value 
relative to halos that lie in a given volume. 
This definition compromises two completely different elements, 
(i) the selection effect and (ii) the effects from individuality of 
the halo mass distribution, the projection and noise.
For our purpose, these elements needs to be examined separately.
We first compute the selection function using
the analytical descriptions of dark matter halos which 
enables us to define ``potentially detectable halos''.
Here, ``potentially'' refers to an ideal case in the absence
of noise.
Then, we shall adopt
the different definition of the {\it completeness} 
that is the fraction of actually detected weak lensing halos relative to
the potentially detectable halos.

Our results have important implications to real observations.
When making an observational strategy, one would like to compute the 
expected number of detectable clusters using, for example, a 
simplified model, and also needs to estimate the expected 
completeness and false detection rate.
Our theoretical model of the weak lensing halo counts may be useful 
for the former case, and our results from the mock data are directly 
applicable to derive the latter estimates.
Also, from the observational point of view, we adopt a directly observable 
quantity, a peak height of weak lensing mass map divided by the noise 
root-mean-square (RMS), that is the signal-to-noise ratio, S/N, 
of the detection, as the fundamental estimate of the lensing signal.

The rest of this paper is organized as follows.
In \S 2, we summarize lensing properties of massive halos 
assuming the universal density profile proposed by Navarro, Frenk \& 
White (Navarro, Frenk \& White 1996; 1997).
Using the expected properties, we compute the selection function
of the weak lensing cluster survey.
Then, we combine the detectability of weak lensing halos with the 
halo mass function based on the Press-Schechter approach 
(Press \& Schechter 1974; we adopt the modified fitting function by 
Sheth \& Tormen 1999) to compute the
weak lensing halo counts and the selection functions with respect to 
the halo mass and redshift.
This simple analytical approach should offer a 
theoretical basis to explore properties of weak lensing halos.
In \S 3, we describe our numerical experiment of weak lensing surveys 
and weak lensing ray-tracing technique. 
Statistical properties of halo catalogs are obtained 
and compared with the theoretical predictions derived in \S 2.
Using the mock weak lensing survey data, in \S 4, we examine the number 
counts of peaks in the weak lensing convergence map, and the theoretical 
prediction developed in \S 2 is tested.
In \S 5, we investigate correspondences between halos in the halo catalog 
and peaks identified in the lensing convergence map.
In particular, we examine the completeness and efficiency of the weak 
lensing halo search paying special attention to the points (1-3) above.
Summary and discussion are given in \S 6.
In Appendix A, we present some lensing convergence maps 
in which a missing halo or a false peak exists for
illustrative examples of irregular systems.
In Appendix B, we develop a correction scheme to the theoretical
prediction of the weak lensing peak counts.
In Appendix C, relations between the halo shape (and orientation) and 
the strength of lensing signal are examined.

Throughout this paper, we work with the standard flat $\Lambda$CDM 
cosmological model with the density parameter $\Omega_{\rm m}=0.3$, 
the cosmological constant $\Omega_\Lambda=0.7$, and Hubble constant 
$H_0=100h~ {\rm km~ s}^{-1}{\rm Mpc}^{-1}$ with $h=0.7$.
We adopt the fitting function of the CDM power spectrum by Bardeen 
et al.~(1986) with the normalization of $\sigma_8=0.9$. 
We take the observational parameters of a typical weak lensing survey 
from a ground based $4-10$m telescope at the present.
To be specific, we consider an imaging observation with a limiting 
magnitude fainter than $R=25.5$ mag in a sub-arcsecond seeing condition,
which provides the galaxy number density $n_g \ga 30$ arcmin$^{-2}$ with
the mean redshift $z\simeq 1$.

\section{Predictions from standard models} 
\label{sec:models} 

\subsection{Lensing properties of an NFW halo}
\label{sec:NFWlens}

The universal density profile of dark matter halos proposed by
Navarro, Frenk \& White (1996; 1997; NFW hereafter) is given by
\begin{equation}
\label{eq:rho-nfw}
\rho_{\rm nfw}(x)={{\rho_s} \over {x (1+x)^2}}, \quad x={r\over{r_s}},
\end{equation}
where $r_s$ is the scale radius.
It is convenient to introduce the concentration parameter 
$\cnfw=\rvir/r_s$ where $\rvir$ is the virial radius.
Bullock et al. (2001) found from $N$-body simulations that the concentration
parameter is related to halo mass as
\begin{equation}
\label{cnfw}
\cnfw(M,z)={{c_\ast}\over {1+z}}
\left( {{M}\over {10^{14}h^{-1}M_\odot}}\right)^{-0.13}, 
\end{equation}
where we set $c_\ast = 8$ for the $\Lambda$CDM model.
It is important to notice that there is a relatively large scatter 
in this relation (Jing 2000). 
The mass enclosed within a sphere of radius $\rvir$ gives the virial
mass by definition:
\begin{equation}
\label{nfw:Mvir}
M_{\rm vir} = {{4\pi \rho_s \rvir^3} \over {\cnfw^3}}
\left[\log(1+\cnfw)-{\cnfw\over{1+\cnfw}}\right]. 
\end{equation}
The virial mass is also defined by the spherical top-hat collapse model
as
\begin{equation}
\label{stm}
M_{\rm vir}={{4\pi}\over 3} \delta_{\rm vir}(z) \bar{\rho}_0 \rvir^3,
\end{equation}
where $\delta_{\rm vir}$ is the threshold
over-density for spherical collapse (see Nakamura \& Suto 1997 and Henry
2000 for useful fitting functions).
{}From equations (\ref{nfw:Mvir}) and (\ref{stm})
one can express 
$\rho_s$ in terms of $\delta_{\rm vir}(z)$ and $\cnfw$.
Introducing $\delta_s=\rho_s/\bar{\rho}_0-1$, one finds,
\begin{equation}
\label{delta_s}
\delta_s={{\delta_{\rm vir}} \over 3} {{\cnfw^3} \over 
{\log(1+\cnfw)-\cnfw/(1+\cnfw)}}.
\end{equation}

Although the NFW profile formally extends to the infinity, an actual 
halo has a finite extent.
We consider a truncated NFW halo, the density profile of which is 
truncated at the virial radius.
The surface mass density of the truncated NFW halo
is given by the line-of-sight integration of the density profile 
(Takada \& Jain 2003):
\begin{equation}
\label{nfw:Sigma}
\Sigma(y)=\int_{-\sqrt{\cnfw^2-y^2}}^{\sqrt{\cnfw^2-y^2}} 
dz~\rho_{\rm nfw}({\bm{y}},z) =  
2 \rho_s r_s f(y),\quad y={r\over{r_s}},
\end{equation}
with
\begin{equation}
\label{f}
f(y)=\left\{
\begin{array}{l}
-{{\sqrt{\cnfw^2-y^2}} \over {(1-y^2)(1+\cnfw)}}
+{1 \over (1-y^2)^{3/2}} {\rm arccosh} {{y^2+\cnfw} \over {y(1+\cnfw)}},\\
\hspace{16em} \mbox{for $y<1$}\\
{{\sqrt{\cnfw^2-1}} \over {3(1+\cnfw)}} 
\left( 1+ {1 \over {1+\cnfw}} \right),
\hspace{2em} \mbox{for $y=1$},\\
-{{\sqrt{\cnfw^2-y^2}} \over {(1-y^2)(1+\cnfw)}}
-{1 \over (y^2-1)^{3/2}} {\rm arccos} {{y^2+\cnfw} \over {y(1+\cnfw)}},\\
\hspace{13em} \mbox{for $1<y \le c$},\\
0, \hspace{2em} \mbox{for $y > c$}.
\end{array}
\right.
\end{equation}
Note that the above expression differs from that by Bartelmann (1996) 
and Wright \& Brainerd (2000) which is for the non-truncated NFW halo
and gives a larger surface mass density than ours.
The lensing convergence profile is given by
\begin{equation}
\label{kappa-nfw}
\kappa(y)={{\Sigma(y)}\over {\Sigma_{cr}}}
= \kappa_s f(y),
\end{equation}
with the critical surface mass density 
\begin{equation}
\label{sigmacr}
\Sigma_{cr}={{c^2}\over{4 \pi G}}{{D_s}\over {D_l D_{ls}}},
\end{equation}
and 
\begin{equation}
\label{kappas}
\kappa_s=
3 \Omega_{\rm m} \delta_s r_s
\left( {{H_0} \over c} \right)^2
{\chi_l (\chi_s-\chi_l) \over {a(z_l) \chi_s}},
\end{equation}
where $a$ is the scale factor normalized at the present time, 
$D$ denotes the angular diameter distance, $\chi$ denotes the comoving 
radial distance (which is equal to the comoving angular diameter distance
for a flat universe), and the subscripts $l$ and $s$ stand for
lens and source, respectively.
It is worth noting that $\kappa$ scales with $\cnfw$ approximately as
$\kappa \propto \cnfw$ for $1\la \cnfw\la 5$ and
$\kappa \propto \cnfw^{1.5}$ for $5\la \cnfw\la 30$.
This clearly proves that the convergence amplitude is sensitive 
to the halo concentration.

\begin{figure}
\begin{center}
\begin{minipage}{8.4cm}
\epsfxsize=8.4cm 
\epsffile{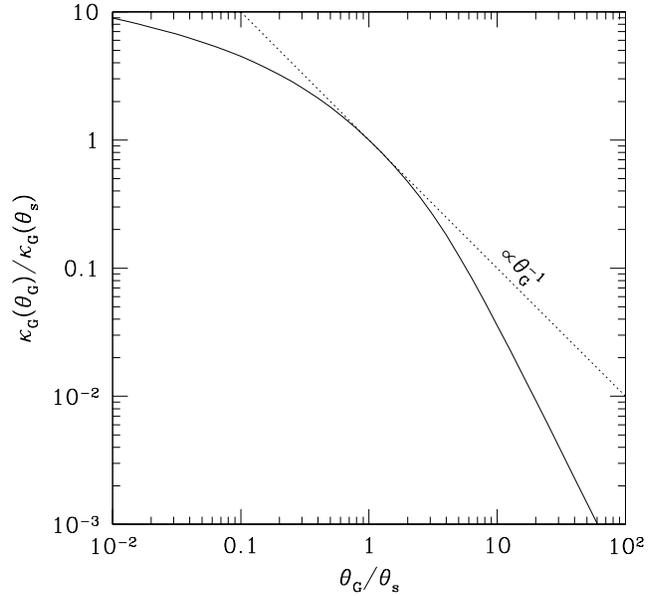}
\end{minipage}
\end{center}
\caption{The solid line shows the lensing signal from an NFW halo smoothed 
with the Gaussian kernel (eq.~[\ref{kappag}]) against the smoothing length. 
For the concentration parameter, $\cnfw=5$ is adopted here.
The dotted line shows scaling of the noise RMS with the smoothing scale 
($\sigma_{\rm noise}\propto \theta_G^{-1}$, see eq.~[\ref{sigma-noise}]).}  
\label{fig:kappag}
\end{figure}

For the number density of source galaxies and the 
intrinsic ellipticity RMS we are interested in, 
a raw (un-smoothed) convergence map is very noisy.
In order to have a lensing mass map with reasonably small amplitude of 
the noise, we smooth the convergence map with a Gaussian window function,
\begin{equation}
\label{WG}
W_G(x)={1\over{\pi \theta_G^2}} \exp\left(-{{x^2}\over{\theta_G^2}}\right).
\end{equation}
The Gaussian smoothing is indeed 
adopted in actual weak lensing surveys (e.g., Miyazaki et al.~2002).
As shown in Van Waerbeke (2000), if the intrinsic ellipticities are
uncorrelated between different source galaxies, 
statistical properties of the resulting noise field can be 
described by the Gaussian random field
theory (Bardeen et al. 1986; Bond \& Efstathiou 1987), on scales where
the discreteness effect of source galaxies can be ignored. 
The Gaussian field is specified by the variance of the noise field that
is specified by the number of galaxies contained within a smoothing aperture
(Kaiser \& Squires~1993; Van Waerbeke 2000):
\begin{equation}
\label{sigma-noise}
\sigma_{\rm noise}^2 = {{\sigma_\epsilon^2}\over 2} 
{1 \over {2\pi \theta_G^2 n_g}},
\end{equation}
where $\sigma_\epsilon$ is the RMS amplitude of the intrinsic ellipticity 
distribution and $n_g$ is the number density of source galaxies.
Throughout this paper we adopt $\sigma_\epsilon=0.4$ and $n_g=30$ 
arcmin$^{-2}$, which are typical values for a  
survey with a limiting magnitude fainter than $R=25.5$ mag in a sub-arcsecond 
seeing condition.
With this choice, the noise RMS 
$\sigma_{\rm noise}=0.02 \times (1$ acrmin$/\theta_G)$.
It is worth noting that $\sigma_{\rm noise}$ is a direct observable in
sense that it can be estimated from a reconstructed weak lensing mass 
map obtained after randomizing the orientation of source galaxy images.

\begin{figure}
\begin{center}
\begin{minipage}{8.4cm}
\epsfxsize=8.4cm 
\epsffile{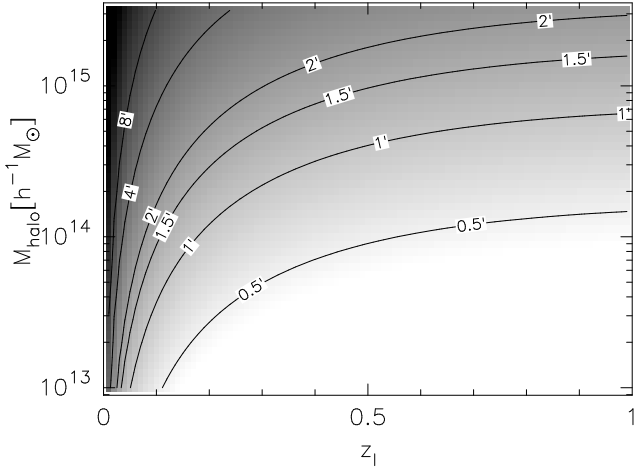}
\end{minipage}
\end{center}
\caption{The gray scale with contour lines show the angular 
scale radius $\theta_s$ in unit of arcminutes, in the halo mass 
($M_{\rm halo}$) and redshift ($z_l$) plane. 
The source redshift is fixed to be $z_s=1$.}  
\label{fig:mass-z_vs_ts}
\end{figure}

We need to also take into account the smoothing effect on the lensing signal 
from a halo. 
The peak amplitude of the convergence profile of halos is modified by 
smoothing into
\begin{equation}
\label{kappag}
\kappa_G(\theta_G)=\int d^2\phi W_G(\phi;\theta_G)\kappa(\phi),
\end{equation}
where the center of the smoothing kernel is set to the 
halo center. Hereafter we refer to this peak amplitude as ``the
lensing signal due to a halo''. 
Figure \ref{fig:kappag} plots the lensing signal as a
function of the smoothing scale.
The dotted line shows a scaling of the noise RMS
with the smoothing scale ($\sigma_{\rm noise}\propto \theta_G^{-1}$).
The comparison between the solid and dotted curves thus shows how 
the signal-to-noise ratio (S/N) varies with the smoothing scale. 
Clearly, an optimal S/N can be attained for 
$\theta_G=(1 - 2)\times \theta_s$.  
Figure \ref{fig:mass-z_vs_ts} shows the contour plot of $\theta_s$ in the 
halo mass and redshift plane, where the mean mass-concentration relation
given by  eq.~(\ref{cnfw}) is employed.
Note that throughout this paper, the source redshift is fixed to be
$z_s=1$, which is a typical value for the mean redshift of source galaxies
for an actual 
survey with limiting magnitude $R\simeq 25.5$.
Combining Figures \ref{fig:kappag} and \ref{fig:mass-z_vs_ts}, 
it is suggested that $\theta_G\sim 1$ arcmin would be 
an optimal choice for efficient survey of massive halos with 
$M_{\rm halo}\ga 10^{14}h^{-1}M_\odot$ at intermediate redshifts.

\begin{figure}
\begin{center}
\begin{minipage}{8.4cm}
\epsfxsize=8.4cm 
\epsffile{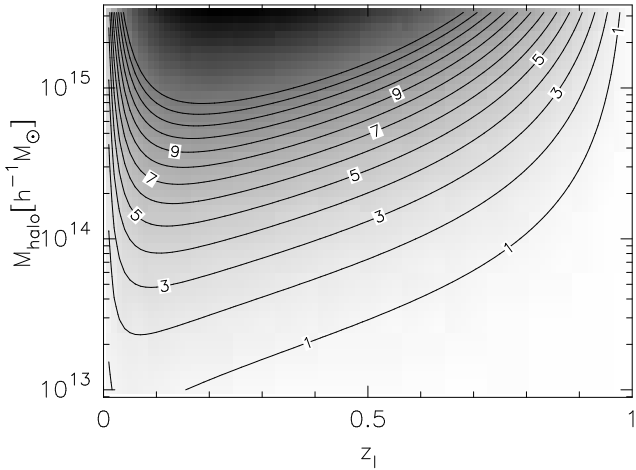}
\end{minipage}
\end{center}
\caption{The gray scale with contour lines shows the S/N value for weak 
lensing halo detection defined by eq.~(\ref{nu}).
The S/N value is computed assuming the universal NFW profile.
We adopt, throughout this paper, $\sigma_\epsilon=0.4$, $n_g=30$arcmin$^{-1}$ 
and $\theta_G=1$ arcmin.}
\label{fig:mass-z_vs_nu}
\end{figure}

We define the signal-to-noise ratio, S/N, for a weak lensing halo detection 
by\footnote{Note on the notation issue: Throughout this paper, we denote 
the S/N of weak lensing signals by $\nu$ and distinguish its type by
the subscript; 
$\nunfw$ stands for the S/N computed assuming the universal NFW 
profile, while $\nulens$ and $\nunoisy$ for the S/N measured 
in the noise-free and noisy realizations of numerical weak lensing mock 
data, respectively
(see \S \ref{sec:simulations} and \S \ref{sec:result-1}). We will also use
the notations $\nuth$ and $\numin$ to denote
a threshold value of the detection and the minimum value to be set
(see \S \ref{sec:result-1} and \S \ref{sec:result-2}), respectively.}
\begin{equation}
\label{nu}
\nu={{\kappa_G}\over {\sigma_{\rm noise}}}.
\end{equation}
Figure \ref{fig:mass-z_vs_nu} shows the contour plot of the S/N value 
computed assuming the universal NFW profile in the halo mass and redshift 
plane.
It is evidently seen from this figure that the lensing signal not only 
depends on the mass but also on the redshift.
This redshift dependence enters mainly through the lensing efficiency 
function $D_l D_{ls}/D_s$ (see eqs.~[\ref{sigmacr}] and [\ref{kappas}]).
This makes the mass sensitivity of weak lensing surveys strongly
redshift dependent.
In other words, the lower mass limit of the detectable halos depends 
strongly on the redshift.
This figure also implies that the {\it effective survey volume} of
a weak lensing halo search depends on the minimum halo 
mass that one aims at locating.
 
It should be noticed that we adopt the {\it mean} 
mass-concentration relation to compute the model prediction for the 
S/N value, though the relation has large scatters
(Bullock et al.~2001; Jing 2000).
Since the lensing signal, $\kappa_G$, depends on 
the concentration parameter as $\kappa_G\propto \cnfw^{1-1.5}$,
the scatter likely translates to scatter in the $\kappa_G$ value
or equivalently in the S/N. 
It should be also noted that the lensing signal for a realistic halo
is affected by
substructures and asphericity of the mass distribution seen in
$N$-body simulations. We will carefully investigate 
these effects, using mock data of weak lensing survey. 

\subsection{Model prediction of weak lensing halo counts}
\label{sec:modelcounts}

\begin{figure}
\begin{center}
\begin{minipage}{8.4cm}
\epsfxsize=8.4cm 
\epsffile{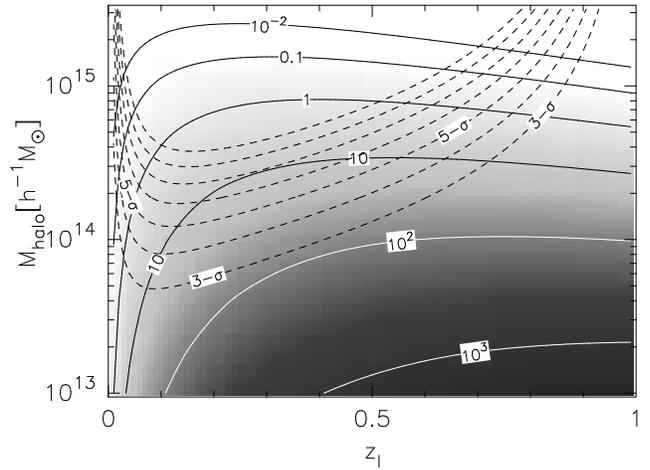}
\end{minipage}
\end{center}
\caption{The gray scale with solid contour lines shows the number
density of halos per 1 deg$^2$ area, a given redshift interval $dz$ and 
a given logarithmic mass interval $d\log M$: 
$d^2N_{\rm halo}/dz/d\log M \equiv dV/dz~dn_{\rm halo}/d\log M$.
This is the integrand of eq.~(\ref{intcounts}).
The dashed contours show the {\it expected} peak height $\nunfw$ 
from 3 to 9 at an interval of $\Delta\nunfw=1$.
Integrating this number over the peak height above a threshold value $\nuth$ 
gives the expected number of peaks per 
1 deg$^2$ as given by eq.~(\ref{intcounts}).
}
\label{fig:dn_dzdlogm}
\end{figure}

\begin{table}
\caption{Summary of the weak lensing peak counts computed using 
eq.~(\ref{intcounts}) for various smoothing scales.
Observational parameters adopted are $\sigma_\epsilon=0.4$, 
$n_g=30$arcmin$^{-1}$, and a fixed source redshift of $z_s=1$.
The second and third columns are for the counts $\nu>5$ and $\nu>4$,
respectively.}
\label{table:intcounts}
\begin{tabular}{lcc} 
\hline
$\theta_G$ [arcmin] & $N(\nu >5)$ [1sq.~deg$^{-1}$] & 
$N(\nu >4)$ [1sq.~deg$^{-1}$]\\
\hline
0.5 & 1.47 & 3.38 \\
1 & 2.08 & 4.05 \\ 
2 & 1.76 & 3.15 \\
3 & 1.30 & 2.27 \\
4 & 0.97 & 1.67 \\
\hline
\end{tabular}
\end{table}

\begin{table}
\caption{Summary of the angular number density (per 1 deg$^2$)
of halos with mass 
more massive than a given minimum halo mass  
(the first column). 
The second and third columns are for the maximum redshifts
$z<1$ and $z<0.7$, respectively.}
\label{table:halocounts}
\begin{tabular}{lcc}
\hline
Halo mass [$h^{-1}M_\odot$] & 
\multicolumn{2}{c}{$N_{\rm halo}$ [1sq.~deg$^{-1}$]} \\
{} & $z<1$ & $z<0.7$ \\ 
\hline
$M>5\times 10^{13}$ & 55 & 30\\
$M>1\times 10^{14}$ & 17 & 10\\
$M>2\times 10^{14}$ & 4.1 &  2.7\\
$M>5\times 10^{14}$ & 0.36 & 0.28\\
\hline
\end{tabular}
\end{table}

Halos that induce a sufficiently high lensing signal are 
likely to be identified as peaks in the convergence map,
even if the map is 
contaminated by the noise due to intrinsic ellipticities. Hence,
it is convenient to work with the number counts of peaks above a certain 
threshold value: $N(\nu>\nuth;\theta_G)$.
In order to compute the number counts, 
we assume that a high peak is generated by the lensing signal from a single 
massive halo and that the number of false peaks due to noise 
(i.e., contamination) 
is much smaller than that of real peaks, as will be shown below in detail.
For our fiducial choice of the smoothing
scale $\theta_G=1$ arcmin,
this assumption is reasonably valid for sufficiently high peaks
(say $\nu>4$), and thus we focus on such high peaks.
In this case, the number density of peaks with height greater than
a threshold $\nu_{\rm th}$ per unit solid angle
is given by (Kruse \& Schneider 2000; 
Bartelmann, King \& Schneider 2001)
\begin{eqnarray}
\label{intcounts}
N(\nu>\nuth;\theta_G,z_s)&=&\int d\chi~{{dV}\over {d \chi}}
 \int dM~{{dn_{\rm halo}(M,z(\chi))}\over {dM}} \nonumber\\
&&\times{\cal{H}}(\nu(M,z)-\nu_{\rm th}),
\end{eqnarray}
where $\chi$ is the radial comoving distance, $dV/d\chi$ is the comoving 
volume element per unit solid angle ($dV/d\chi =\chi^2$ for a flat
cosmology), $dn_{\rm halo}/dM$ is the halo mass function, for which 
we adopt the fitting function by Sheth \& Tormen (1999),  
and ${\cal{H}}(x)$ is the Heaviside step function
(${\cal{H}}(x)=1$ for $x>0$ and 0 otherwise).
The differential counts are obtained by
\begin{equation}
\label{defcounts}
n_{\rm peak}(\nu) = 
\left| {{dN(\nu>\nuth;\theta_G,z_s)} \over {d\nu}} \right|.
\end{equation}

Table \ref{table:intcounts} summarizes the predicted weak lensing peak counts 
for various smoothing scales.
It is shown that a choice of $\theta_G=1$ arcmin gives the largest counts, for
$\sigma_\epsilon=0.4$ and $n_g=30 $ arcmin$^{-2}$.
In the following discussion,
we will adopt this smoothing scale.

In Figure \ref{fig:dn_dzdlogm}, we plot the selection function of
weak lensing halo search with respect to the halo mass and redshift.
The gray scale with solid line contours shows the number density of
halos per 1 deg$^2$ against a given redshift interval of $dz$ and a
given logarithmic mass interval of $d\log M$:
\begin{equation}
{{d^2N_{\rm halo}}\over {dz~d\log M}}\equiv 
{{dV}\over{dz}} {{dn_{\rm halo}}\over {d\log M}}. 
\end{equation}
Note that this is the integrand of eq.~(\ref{intcounts}).
It is clear that, for a given halo mass, the number density 
rapidly increases with redshift at low redshifts $z \la 0.2$, and
then gradually decreases toward higher redshifts. 
This is because the volume element, $dV\propto \chi^2 d\chi$, 
increases with redshift, whereas the halo number density decreases.
The dashed contours show the expected peak height $\nunfw$ 
for the universal NFW halo at intervals of $\Delta\nunfw=1$. 
The comparison between the solid and dashed contours clarifies 
that, if threshold peak 
height $\nuth=4-5$ is employed, 
the weak lensing halo search is most sensitive to halos that 
are located at the redshift 
between $0.1\la z\la 0.7$ and have masses
greater than $10^{14}h^{-1}M_\odot$.

\begin{figure}
\begin{center}
\begin{minipage}{8.4cm}
\epsfxsize=8.4cm 
\epsffile{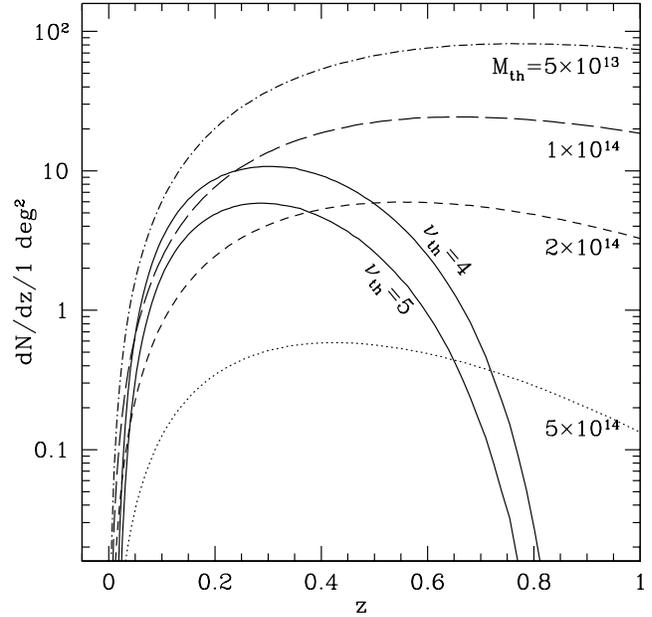}
\end{minipage}
\end{center}
\caption{The solid lines show the redshift distribution of halos with a peak 
height above a threshold value $\nuth$ per 1 deg$^2$, 
$dN(\nunfw>\nuth)/dz/1$ deg$^2$, computed using 
eq.~(\ref{intcounts}).
The upper and lower solid lines are for $\nuth=4$ and 5, respectively.
Note that we employed $\theta_G=1$ arcmin and $z_s=1$. The
broken lines show the same redshift distributions but for halos with 
masses above a given minimum mass, 
$dN_{\rm halo}(M>M_{\rm th})/dz/1$ deg$^2$.}
\label{fig:dndz}
\end{figure}

\begin{figure}
\begin{center}
\begin{minipage}{8.4cm}
\epsfxsize=8.4cm 
\epsffile{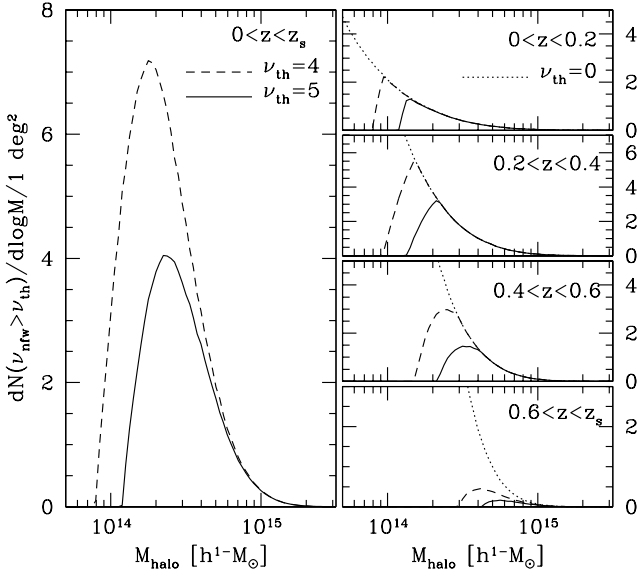}
\end{minipage}
\end{center}
\caption{{\it Left Panel}:  
Mass function of halos with $\nunfw>\nuth$ computed using 
eq.~(\ref{intcounts}).
The threshold values are $\nuth=4$ and 5 for the dashed and solid lines, 
respectively,
The source redshift is taken to be $z_s=1$. 
{\it Right Panels}:  
Same as the left panel but divided into four ranges of the halo redshift 
as denoted in each plot.
The dashed lines show the mass function of halos within the corresponding 
volume (which is equivalent to set $\nuth>0$, that is, no limit on the weak 
lensing signal is imposed).
Note that the scales of y-axis are labeled on the right side of the plots.}
\label{fig:dndm-zrange}
\end{figure}

Let us look further into the redshift- and mass-dependence of 
the weak lensing selected halo counts.
The solid curves in 
Figure \ref{fig:dndz} show the redshift distribution
of halos selected by the weak lensing signal. These curves
can be compared with the distribution of halos above a certain mass
threshold (broken lines).
A remarkable difference between these two is evidently seen at 
redshifts $z>0.5$;  the distribution of weak lensing selected 
halos drops very rapidly, while the mass selected ones decrease very 
gradually.
This is again because the lensing efficiency peaks around $z=0.4$ for
source galaxies of $z=1$. 
Figure \ref{fig:dndm-zrange} shows essentially the 
same information but from a different angle.
The left panel shows the mass function of halos with 
peak height exceeding a threshold value (the cases of $\nuth=4$ and 5 are 
displayed). 
The lower mass limit in the mass function 
is determined by the detection threshold, while the 
decline in the high mass side (which has an approximate exponential slope) 
is due to the property of the halo mass function 
(i.e., the exponential shape of its high mass end, see 
Kruse \& Schneider 2000).
Right panels of Figure \ref{fig:dndm-zrange} shows the mass distribution 
of halos for different redshift intervals as labeled in each plot.
The solid (dashed) lines are for halos with $\nunfw>5$ ($\nunfw>4$), 
while the dotted lines are for all the halos in the corresponding volume.
The upper two plots clearly show that, up to $z=0.4$,
almost all the halos with $M>2\times10^{14}h^{-1}M_\odot$ indeed produce
high peaks with $\nunfw>5$.
However, for higher redshifts, only very massive halos 
can produce high peaks.

To summarize the results shown in 
Figures \ref{fig:dn_dzdlogm} - \ref{fig:dndm-zrange},
we argue that weak lensing is not an efficient technique of finding 
high-redshift halos (say $z>0.6$), except for very massive halos of 
$M>5\times10^{14}h^{-1}M_\odot$,  if a
mean source redshift is around $z_s=1$.
It should be noticed that 
the ability of the weak lensing halo surveys
depends on observational conditions. 
As a survey is deeper, the halo search more enables us to detect
less massive halos and halos at higher redshift.

Table \ref{table:halocounts} gives the number density of halos with
mass above a given minimum mass, which is obtained by integrating the
halo mass function up to a given maximum redshift.  Comparison with
Table \ref{table:intcounts} provides a rough estimate of how weak
lensing survey misses mass-selected halos in the given survey volume.
A large discrepancy between the halo counts and the weak
lensing halo counts is seen, which arises from the redshift dependence
of the lensing selection function as shown in Figures \ref{fig:dndz} and
\ref{fig:dndm-zrange}.

\section{Numerical simulation and mock catalogs}
\label{sec:simulations}

The theoretical model prediction discussed in the previous section provides 
us with a basic selection function of the weak lensing cluster survey.
However, since our model is based on the simple 
averaged descriptions of the dark matter halos, it does not  
take into account the individuality of halos which causes a 
substantial scatter in the lensing signal
and may even cause a bias.
Furthermore, there are two additional sources of scatter; the projection 
effect and the intrinsic ellipticity noise.
In this section, we will examine these effects in detail
using mock numerical experiments.

\subsection{$N$-body simulation}
\label{sec:nbody}

We used $N$-body data from Very Large $N$-body Simulation (VLS) carried 
out by the Virgo Consortium (Jenkins et al.~2001, and see also Yoshida
et al.~2001 for simulation details).
The simulation was performed using a parallel P$^3$M code 
(MacFarland et al.~1998) with a force softening
length of $l_\mathrm{soft}\sim 30\,h^{-1}\mathrm{kpc}$. The simulation
employed $512^3$ CDM particles in a cubic box of
$479\,h^{-1}\mathrm{Mpc}$ on a side. It uses a flat cosmological model
with $\Omega_0=0.3$, $\Omega_\Lambda=0.7$, and $h=0.7$. 
The initial matter power spectrum was computed using CMBFAST (Seljak \&
Zaldarriaga 1996) assuming a baryonic matter density of
$\Omega_\mathrm{b}=0.04$. 
The normalization of the power spectrum is set by $\sigma_8=0.9$.
The particle mass
($m_\mathrm{part}=6.86\times 10^{10}h^{-1}M_\odot$) of the simulation
is sufficiently small to guarantee practically no discreteness effect
on dark-matter clustering on scales down to the softening length in
the redshift range of interest for our purposes (Hamana, Yoshida \&
Suto 2002).

\subsection{Weak lensing ray-tracing simulation}
\label{sec:raytracing}

The multiple-lens plane ray-tracing algorithm we used is detailed in
Hamana \& Mellier (2001; see also Bartelmann \& Schneider 1992, 
Jain, Seljak \& White 2000 and Vale \& White 2003 for the theoretical 
basics and technical issues); here we thus
describe only aspects specific to the VLS $N$-body data. 
In order to generate the density field between $z=0$ and
$z\simeq1$, we use a stack of 6 snapshot outputs from two runs of the
$N$-body simulation, which differ only in the realization of the
initial fluctuation field. Each cubic box is divided into 4 sub-boxes
of $479^2\times 119.75h^{-3}\mathrm{Mpc}^3$ with the shorter box side
being aligned with the line-of-sight direction. $N$-body particles
in each sub-box are projected onto the plane perpendicular to the
shorter box side and thus to the line-of-sight direction. In this way,
the particle distribution between the observer and $z\simeq1$ is
projected onto $19$ lens planes separated equally by
$119.75\,h^{-1}\mathrm{Mpc}$. Note that in order to minimize the
difference in redshift between a lens plane and an output of $N$-body
data, only one half of the outputs (i.e.~two sub-boxes) from $z=0$ $N$-body 
box was used.

The particle distribution on each plane is converted into the surface
density field on a $4096^2$ regular grid using the
triangular shaped cloud (TSC) assignment scheme (Hockney \& Eastwood
1988). $4096^2$ grid is chosen to maintain the resolution provided by
the $N$-body simulation as well as to remove at the same time the shot noise
due to discreteness in the $N$-body simulation. The computation 
follows the procedure described in Hamana \& Mellier (2001) and 
Jain et al.~(2000).

Having produced surface density fields on all lens planes, we trace 
$512^2$ rays backwards from the observer's point to source plane using the
multiple-lens plane algorithm (e.g.~Schneider, Ehlers \& Falco
1992). The initial ray directions are set on $512^2$ grids with a
grid spacing of $0.24\,\mathrm{arcmin}$, and thus the total area covered by
the rays is $2.048^2\,$square degrees. We made 50 realizations of the
underlying density field by randomly shifting the simulation boxes in
the direction perpendicular to the line-of-sight using the periodic
boundary conditions of the simulation\footnote{
The weak lensing ray-tracing simulation data used in this paper are 
available from T.~Hamana on request, hamanatk@cc.nao.ac.jp.}. 
Note that the lens planes coming from the same box are shifted in the 
same way in order to maintain the clustering of matter in one box.

In the following analyses, we use the lensing convergence 
map of the source plane closest to $z=1$.
This is, we call, the {\it noise-free lensing map}.

\subsection{Adding noise}
\label{sec:noise}

To simulate a more realistic lensing survey, we add noises due to
intrinsic ellipticities of source galaxies to the simulated lensing
map. 
Actual observations are further affected by other systematics
such as imperfect conditions of telescope and/or of electric devices 
of detector.
Examining these effects is beyond the scope of the present paper,
and thus we do not consider such observational effects.

As extensively 
shown in Van Waerbeke (2000), the noise field due to the
intrinsic ellipticities is well approximated by the Gaussian random
field on angular scales of our interest, if the ellipticities are
uncorrelated between different sources (see also Jain \& Van Waerbeke
2000). 
The Gaussian noise added to our $\kappa$ map has variance
\begin{equation}
\label{sigma-pix}
\sigma_{pix}^2={{\sigma_\epsilon^2} \over 2}{1 \over {n_g \theta_{pix}^2}},
\end{equation}
where $\theta_{pix}=0.24$ arcmin is the pixel size of our $\kappa$ maps.
The noise RMS on pixels is  $\sigma_{pix}=0.22$ which is much larger than 
the RMS of the lensing convergence field.

\subsection{Halo catalogs on the light-cone}
\label{halocatalog}

\begin{figure}
\begin{center}
\begin{minipage}{8.4cm}
\epsfxsize=8.4cm 
\epsffile{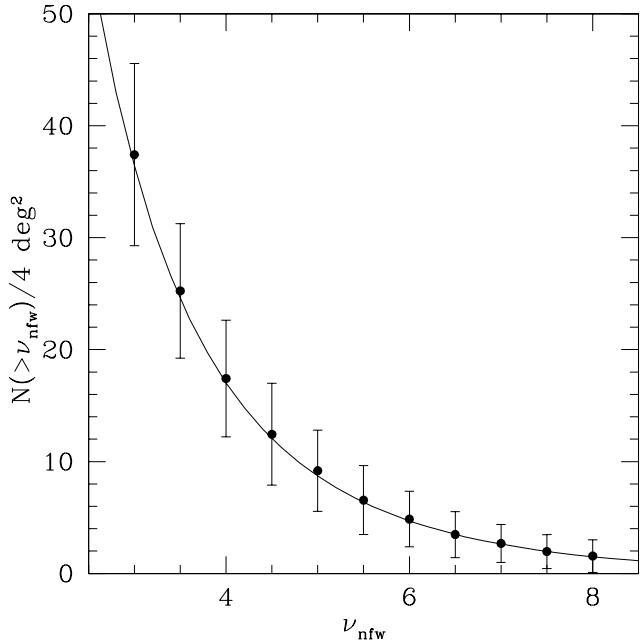}
\end{minipage}
\end{center}
\caption{The number counts of the lensing peaks above
a given threshold $\nunfw$ per a $4$ square degrees area. 
The solid curve shows the model prediction computed from
eq.~(\ref{intcounts}), while the filled circles with error bars denote the
mean and RMS among 50 realizations of mock simulation data.
Note that the peak height of a halo is estimated from the halo mass 
and redshift
in the manner described in the text (\S \ref{halocatalog}).}
\label{fig:counts-halo}
\end{figure}

We identify dark matter halos in 5 low redshift outputs of the $N$-body 
simulation
using the standard friends-of-friends algorithm 
with a linking parameter of $b=0.164$ 
(in units of the mean particle separation).  
Halos with mass greater than $1\times 10^{13} h^{-1}M_{\odot}$, 
which corresponds to the mass of 146 simulation particles, are
used in the analyses below.
We define the halo center to be the center-of-mass 
position of member particles.

\begin{figure}
\begin{center}
\begin{minipage}{8.4cm}
\epsfxsize=8.4cm 
\epsffile{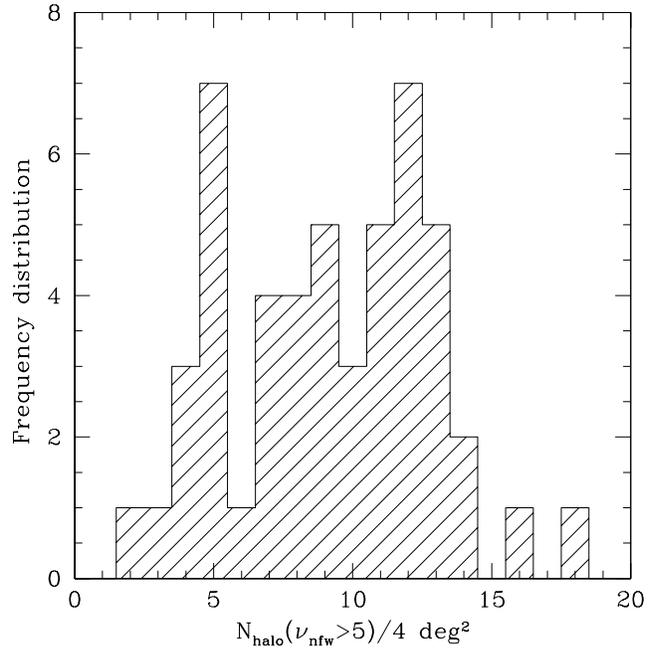}
\end{minipage}
\end{center}
\caption{The frequency distribution of numbers of halos with the expected 
peak height (see \S \ref{halocatalog} for its definition) $\nunfw>5$ from 50 
realizations of 4 square degrees.}
\label{fig:nhalo}
\end{figure}

In order to evaluate the halo shape, 
we compute the inertia tensor of the mass distribution;
\begin{eqnarray}
\label{Iij}
I_{ij}&=&\int d\bm{x}^3 (x_i-\bar{x}_i) (x_j-\bar{x}_j) \rho(\bm{x})\nonumber\\
&=& m_p \sum_{k=1}^{N_p} (x_{i,k}-\bar{x}_i) (x_{j,k}-\bar{x}_j),
\end{eqnarray}
where $i,j$ run from 1 to 3, $\bar{x}_i$ denotes the center-of-mass position
and $N_p$ is the number of member particles of the halo.
{}From the inertia tensor we compute the eigen vectors and convert them
to the axial ratios $a/c$ and $a/b$ ($a\le b\le c$).
We also compute the trace of the inertia tensor,
\begin{equation}
\label{Q}
Q\equiv {1\over 3} (I_{11}+I_{22}+I_{33}).
\end{equation}
This is compared with the expectation value for the NFW halo,
\begin{eqnarray}
\label{Qnfw}
Q_{\rm NFW}&=& {{4\pi} \over 3} \int_0^{r_{\rm vir}} dr~r^4 
\rho_{\rm nfw}(r)\nonumber\\
&=& M_{\rm vir} r_{\rm vir}^2 q(\cnfw),
\end{eqnarray}
with
\begin{equation}
\label{q}
q(\cnfw) ={{\cnfw^3-3\cnfw^2-6\cnfw + 6(1+\cnfw)\log(1+\cnfw)}
\over
{6 \cnfw^2 [(1+\cnfw)\log(1+\cnfw)-\cnfw}]}.
\end{equation}
Note that $q(\cnfw)$ can be approximated by 
$q(\cnfw)\simeq 0.22 \cnfw^{-0.26}$ for $3 \la \cnfw \la 30$.
The ratio $Q/Q_{\rm NFW}$ gives an estimate of the concentration 
of the mass distribution.
We also define the following quantity as an estimate of
the line-of-sight elongation of the halo mass distribution:
\begin{equation}
\label{R}
R ={{2 I_{33}} \over
{I_{11}+I_{22}}},
\end{equation}
where $x_3$ is taken by the direction parallel to the line-of-sight.
Although this is rather a crude estimation, $R>1$ or $R<1$ generally
implies halos elongated along or 
perpendicular to the line-of-sight.

Halo catalogs on the light-cone are generated by 
stacking the simulation 
outputs in the same manner as in the ray-tracing experiments. 
We use the 50 realizations which have the same underlying matter 
distribution as used in the ray-tracing data.
This combined data set allows us to directly relate the distribution of 
dark matter  halos to the weak lensing convergence map.

\begin{figure}
\begin{center}
\begin{minipage}{8.4cm}
\epsfxsize=8.4cm 
\epsffile{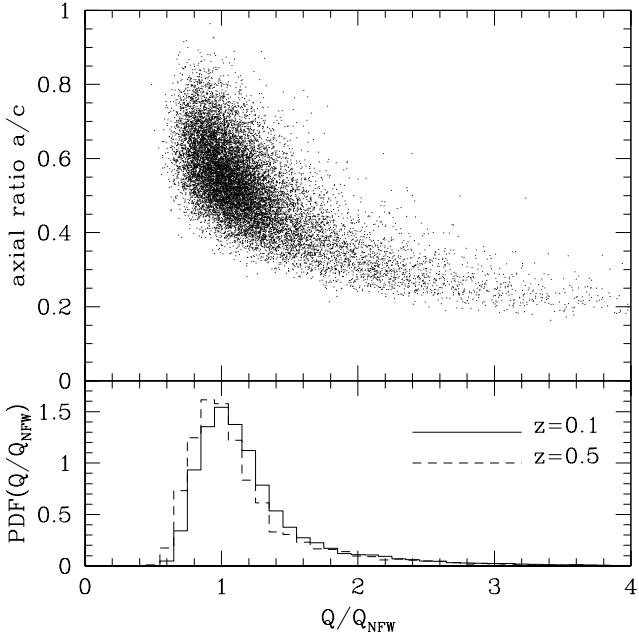}
\end{minipage}
\end{center}
\caption{{\it Upper Panel}: The scatter plot showing the relation
between $Q/Q_{\rm NFW}$ and the axial ratio $a/c$ for halos with 
$M>3\times 10^{13}h^{-1}M_\odot$ in the $z=0.1$ output.
Here $Q_{\rm NFW}$ is defined by eq.~(\ref{Qnfw}) and is computed using
the relation eq.~(\ref{stm}).
{\it Lower Panel}: The PDF of $Q/Q_{\rm NFW}$ of halos with 
$M>3\times 10^{13}h^{-1}M_\odot$, the solid and dashed lines for
$z=0.1$ and $z=0.5$ outputs, respectively.}
\label{fig:Q-ca}
\end{figure}

\begin{figure}
\begin{center}
\begin{minipage}{8.4cm}
\epsfxsize=8.4cm 
\epsffile{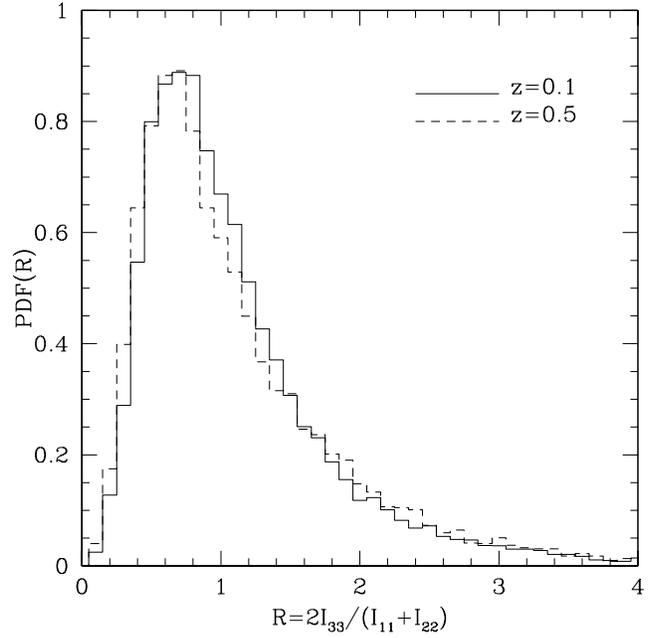}
\end{minipage}
\end{center}
\caption{The PDF of the $R$ defined by eq.~(\ref{R}) for halos with
$M>3\times 10^{13}h^{-1}M_\odot$ at $z=0.1$ (solid line) and 
$z=0.5$ (dashed line) outputs. Note that the median of the distributions is 
$\sim 0.92$, and 55\% of the all samples has $R<1$.}
\label{fig:PDF-R}
\end{figure}

In summary, each halo in the mock catalogs has data on the mass, 
the estimates of the halo shape $Q$ and $R$ defined by eqs.~(\ref{Q}) 
and (\ref{R}), respectively, the redshift computed from the radial 
distance to a halo, and the angular position converted from the spatial 
position. 
In addition, for each halo, 
we compute the virial radius 
and the concentration parameter using eqs.~(\ref{stm}) and (\ref{cnfw}), 
from the halo mass and redshift 
of $N$-body data output time (which are $z=0$, 0.1, 0.3, 0.5 and 0.8). 
These values allow us to evaluate the {\it expected} lensing signal 
using eq.~(\ref{kappag}) with eq.~(\ref{kappa-nfw}), which is correct
only if the halo has the NFW profile.
Similarly, we also compute $Q_{\rm NFW}$ using eq.~(\ref{Qnfw}).
Moreover, we introduce a $flag$ parameter for each halo, to quantify
the following possible problem that arises from 
our use of multiple-lens plane ray-tracing algorithm. 
As a result of projection of $N$-body particles into 4 lens planes, 
member particles of given one halo located close to the boundary could be 
separated into different planes.
Such a split halo would generate an artificially deformed 
(or even separated) lensing signal in a $\kappa$ map.
We assign $flag=1$ for such halos whose virial radius is larger than
the distance between the halo center and the closest plane, 
$flag=2$ for halos whose scale 
radius is larger than the distance to the closest plane, and $flag=0$ 
otherwise.
The fraction of flagged halos is very small, 
0.88\% for $flag=1$ and 0.19\% for $flag=2$, thus it does not
have significant influence on our analyses below, but we carefully check
if flagged halos cause any artificial eccentric signals in the lensing 
convergence map.

The {\it expected} lensing signal of each halo is converted into 
the peak height in the same way as in eq.~(\ref{nu}), 
which we denote as $\nunfw$.
Total numbers of halos with $\nunfw>5$ and $\nunfw>4$ in our 50 realizations 
are 459 and 871, respectively, which are sufficiently large to allow the
statistical analyses of weak lensing halos.

The redshift and mass distributions of halos in the halo catalog 
agree well with the model predictions in \S \ref{sec:modelcounts}.
Figure \ref{fig:counts-halo} plots the accumulate number counts of halos 
against the {\it expected} peak height, $\nunfw$. 
The solid curve is the theoretical prediction of 
eq.~(\ref{intcounts}) and agrees well with the simulation result. 
This agreement is not surprising, because ingredients of the theoretical 
model (the universal profile and halo mass function) are calibrated by
$N$-body simulations. 

The counts of halos with $\nunfw>5$ in $2.048^2$ square degrees is found 
to be $N(\nunfw>5)=9.18\pm 3.63$, where the error represents the RMS among 
50 realizations.
This is in good agreement with the theoretical prediction of 8.71.
However, the RMS is somewhat large and
slightly larger than the Poisson fluctuation. 
Figure \ref{fig:nhalo} shows the distribution of numbers of halos with 
$\nunfw>5$ in each realization, quite a wide spread is evidently seen.
This is mainly due to the very strong clustering of massive halos.
Accordingly, given a limited survey area, the halo counts may 
significantly vary from survey to survey.

\begin{figure*}
\begin{center}
\begin{minipage}{17.7cm}
\epsfxsize=17.7cm 
\epsffile{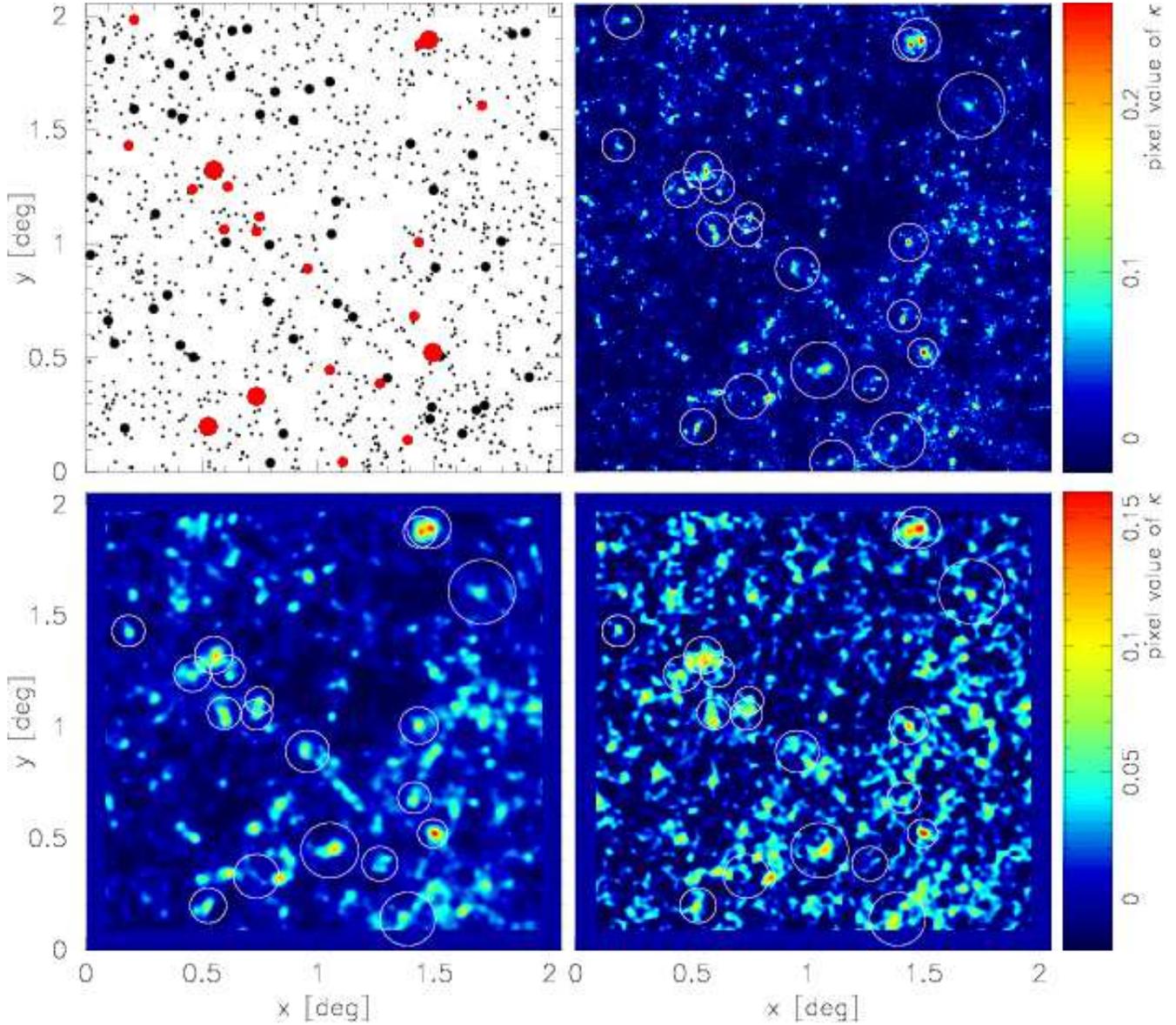}
\end{minipage}
\end{center}
\caption{{\it Upper-left}:
The celestial distribution of halos identified from one realization
of mock light-cone data generated stacking $N$-body outputs.
The large filled circles denote halos with 
$M_{\rm halo}$[$h^{-1}M_\odot$]$>3\times10^{14}$, the small filled circles are 
for $3\times10^{14}>M_{\rm halo}$[$h^{-1}M_\odot$]$>8\times10^{13}$
and the dots are for 
$8\times10^{13}>M_{\rm halo}$[$h^{-1}M_\odot$]$>1\times10^{13}$.
Only the halos within redshift interval between 0.1 and 0.7 are displayed.
{\it Upper-right}:
The gray map shows the lensing convergence map on the $512^2$ pixels 
with the pixel size of 0.24 arcmin obtained from ray-tracing experiment
through the same mass distribution as used in the upper-left panel.
The circles show the positions of halos with $\nunfw>4$, 
the center of which is set at the position of each halo 
and the radius of which shows the projected virial radius 
(see \S. \ref{halocatalog} for details).
{\it Bottom-left}:
Same as the top-left panel but the convergence map 
is smoothed by the Gaussian window function with $\theta_G=1$ arcmin.
The edge regions of the field with 5 arcmin width are masked because 
of incomplete smoothing due to the lack of $\kappa$ data on outer region.
The gray scale of this plot is same as that of the bottom-right panel.
{\it Bottom-right}:
Same as the bottom-left panel but 
the Gaussian noise is added before smoothing is applied.
{}From comparison with the other plots, it is clear that high 
peaks due to massive halos are still identifiable.}
\label{fig:example2x2}
\end{figure*}

The lower panel of Figure \ref{fig:Q-ca} shows the PDFs of $Q/Q_{\rm NFW}$ 
computed from massive halos with $M > 3\times 10^{13}h^{-1}M_\odot$
($Q_{\rm NFW}$ is computed using the mean mass--concentration 
relation eq.~[\ref{cnfw}]).
The PDF peaks at $Q/Q_{\rm NFW}\sim 1$ as expected, but has rather broad 
distribution.
No significant difference in the PDFs between the five different redshift 
boxes was found.
The upper panel shows the scatter plot in $Q/Q_{\rm NFW}$--$a/c$ plane for
halos with $M > 3\times 10^{13}h^{-1}M_\odot$ from $z=0.1$ output.
It is clear that the mass concentration defined by
$Q/Q_{\rm NFW}$ correlates with the axial ratio, which implies that an 
extended mass distribution
is likely associated with eccentric mass distributions such as halos
having substructures or on-going merger.
We note that the PDF of axial ratio peaks at $a/c\sim 0.5$ and thus 
most of halos significantly deviate from the spherical symmetry.
This broad distribution leads to significant scatter in the relation
between the expected and observed lensing signals as seen below.

Figure \ref{fig:PDF-R} shows the PDF of the shape parameter $R$ defined 
by eq.~(\ref{R}) for massive halos with $M > 3\times 10^{13}h^{-1}M_\odot$.
A broad distribution is evidently seen and we checked that 
the distribution does not largely change over the redshift range 
we considered ($z<0.8$).
It is important to note that the median of the distribution is less 
than unity ($\sim 0.92$), thus more than half of the massive halos 
(specifically 55\% for halos with $M > 3\times 10^{13}h^{-1}M_\odot$
irrespective of the redshift) 
have $R<1$.
This is a natural consequence of the random orientation of the halos'
major axis; 
among the spatial three directions, it is more likely that
the major axis of a halo is
elongated along either of two directions perpendicular
to the line-of-sight rather than along it. 
The skewed distribution leads to systematic bias in the relation
between the expected and observed lensing signals
(see \S \ref{sec:halo-lens} and Appendix \ref{sec:haloshape} 
for details).

\subsection{Visual impressions}
\label{visual}

Figure \ref{fig:example2x2} represents 
the lensing convergence maps and the celestial distribution of halos,
which are 
taken from one of the 50 realizations. 
In the top-right panel, the gray scale shows $\kappa$ value on the 
$512^2$ pixels with the pixel size of 0.24 arcmin.
The bottom-left panel shows the same convergence map as in the 
top-right panel but after smoothed by the Gaussian filter function of
$\theta_G=1$ arcmin. Adding the Gaussian noise yields the noisy map 
shown in the bottom-right panel. While
the noise indeed alters the appearance of the convergence map, the 
very high peaks remain identifiable.

The top-left panel shows the distribution of all the halos
identified by the FOF scheme in the original $N$-body simulations from which
the convergence maps in the other plots are generated. 
Note that the size of point scales with halo masses. 
A good correspondence in positions between massive halos and high 
lensing signals on the $\kappa$ map is seen.
This correspondence is more clearly seen if halos are selected by means
of the expected $\nunfw$ value: 
in the convergence maps (the top-right and two bottom panels), the positions 
of halos with $\nunfw>4$ are marked by the circles with their centers 
being placed at the halo center-of-mass position, and their radii show
the angular virial radius (see \S \ref{halocatalog}).
The improvement in the correspondence reflects the fact that the lensing
signal is determined by combined contributions of the halo mass, the
redshift and the halo concentration, rather by the mass alone.

\section{Results on peak counts} 
\label{sec:result-1}

In this section, we study the number
counts of high peaks in the lensing convergence map.
To do this, we use three kinds of $\kappa$ maps;
(i) the noise-free $\kappa$ maps, 
(ii) the noisy $\kappa$ maps where the Gaussian noise was added 
in the manner described in \S \ref{sec:noise},
and (iii) the pure noise maps without the cosmological lensing
signals.
All these maps are smoothed with a Gaussian window function 
with $\theta_G=1$ arcmin (eq.~[\ref{WG}]). 
We identify positive (negative) curvature peaks in the maps 
as pixels that have higher (lower) values of 
$\kappa$ than all the surrounding eight pixels (Jain \& 
Van Waerbeke 2000, and also see Miyazaki et al 2002 for the first
application to actual observational data).
The peak height is defined as the ratio of the convergence value at the
peak to the noise RMS, as given by eq.~(\ref{nu}).
For all the three cases, $\sigma_{\rm noise}=0.02$ ($\sigma_\epsilon=0.4$ 
and $n_g=30$arcmin$^{-2}$) is adopted.

\subsection{The peak PDF} 
\label{sec:peakpdf}

The probability distribution function (PDF) of peaks 
in the convergence map is studied in detail by Jain \& Van Waerbeke (2000).
Here we briefly summarize the characteristics of the peak PDF that 
are relevant for the subsequent discussion.

\begin{figure}
\begin{center}
\begin{minipage}{8.4cm}
\epsfxsize=8.4cm 
\epsffile{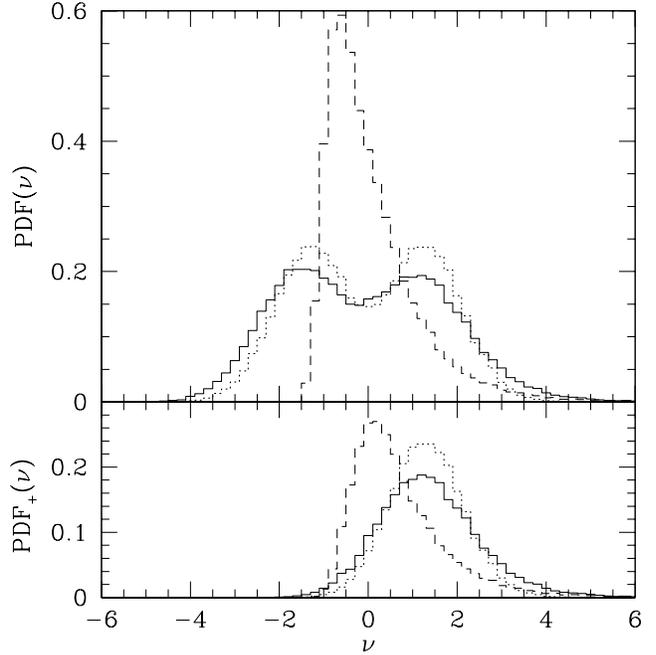}
\end{minipage}
\end{center}
\caption{The simulation results for the PDFs of the peaks in the noise-free 
$\kappa$ (dashed histogram), in the noisy $\kappa$ map (solid histogram) and 
in the pure noise map (dotted histogram).
Shown is the cumulative results from the 50 realizations.
The upper panel shows the sum of the positive and negative curvature peaks,
while the lower panel shows the contribution from the positive peaks only.
In the total $50\times 2.048^2$ square degrees' area, 
there are 59,908, 107,165 and 114,250 peaks for the noise-free 
$\kappa$, the noisy $\kappa$ map and the pure noise map, respectively.
The corresponding numbers for positive curvature peaks are
31,043, 53,123, and 57,141.}
\label{fig:peakpdf}
\end{figure}

Figure \ref{fig:peakpdf} shows the peak PDFs in the noise-free 
$\kappa$ (dashed histogram), the noisy  $\kappa$ map (solid histogram) and 
the pure noise map (dotted histogram).
The upper panel shows the sum of the positive and negative curvature peaks,
while the lower panel shows the contribution from the positive peaks only.
In the absence of the noise, the peak PDF reflects
the characteristics of the nonlinear gravitational clustering. 
Most of negative peaks are due to the negative curvature peaks (troughs).
There is a cutoff in the negative peak height because
the mass density contrast cannot be lower than $-1$. 
One can also find a tail toward the 
high positive peak height which arises from the lensing signals due to
massive halos (Kruse \& Schneider 2000).
Another noticeable feature seen from the lower panel is that some of the 
{\it positive} curvature peaks have a {\it negative} peak height, 
which occurs when the peak due to a halo is superposed on a void region
in the same line-of-sight.
If the noise field is Gaussian, 
the PDF from the pure-noise map is derivable from the 
the Gaussian random field theory 
(Bardeen et al.~1986; Bond \& Efstathiou 1987; Van Waerbeke 2000).
The symmetric bimodal feature of the peak PDF is due to the 
superposition of two contributions from peaks with 
positive and negative curvatures.
While adding the noise to the cosmological
$\kappa$ map alters the peak PDF 
drastically, there remain some characteristic features that contain
useful cosmological information.
First, the asymmetric double peaks are due to the skewed feature of the 
cosmological $\kappa $ PDF as stated above. Second, there are 
excesses over the noise PDF on both high positive 
($\nu>3$) and negative peaks ($\nu<-3$). These features were indeed 
detected in the recent weak lensing 
survey by Miyazaki et al. (2002).

It is worth pointing out that the pure noise PDF can be directly estimated
from an actual observed data. First, we can generate a
pure noise map by applying mass-reconstruction to a randomized catalog
where the orientation of source galaxy images is randomized
(to remove coherent image distortions from lensing signals).
Then we can measure the peak PDF from the pure noise map, 
as was performed in Miyazaki et al.~(2002).
This allows one to estimate the contamination rate 
(the false peak rate) directly from the data.

\subsection{High peak counts} 
\label{sec:peakcounts}

\begin{figure}
\begin{center}
\begin{minipage}{8.4cm}
\epsfxsize=8.4cm 
\epsffile{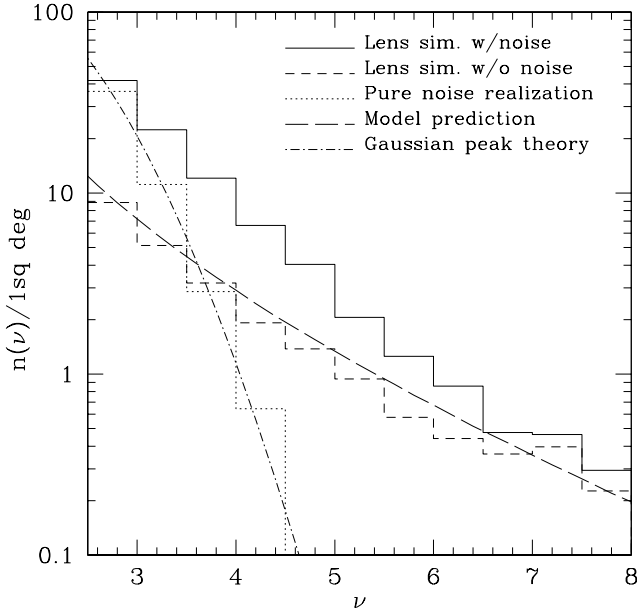}
\end{minipage}
\end{center}
\caption{Counts of peaks per 1 square degrees.
The histograms are from the numerical experiments; 
the noisy $\kappa$ maps (solid),
the noise-free $\kappa$ maps (dashed) and the pure noise realizations (dotted).
The long-dashed line shows the model prediction, eq.~(\ref{defcounts}) with
eq.~(\ref{intcounts}), and the dot-dashed line shows the prediction by 
Gaussian random field theory (e.g., Bond \& Efstathiou 1987; Van Waerbeke 
2000).}
\label{fig:peakcounts}
\end{figure}

In the following, we restrict our discussion to the number counts of
high peaks that are mainly due to massive halos.  
Figure \ref{fig:peakcounts} shows the number density
of (positive curvature) peaks per 1 square degrees.
The histograms show the average values computed from 
the 50 realizations of the numerical experiments;
the noisy $\kappa$ maps (solid), the noise-free $\kappa$ maps (dashed) 
and the pure noise maps (dotted).
The long-dashed line shows the model prediction from 
eq.~(\ref{defcounts}),
while the dot-dashed line shows the prediction for
the pure noise counts (Van Waerbeke 2000). 
{}From comparison of the results for the lensing maps with and
without the noise, it is clear that the noise 
significantly increases the peak counts over a
range of peak heights we have considered.
Even high peaks with $\nu>5$ are similarly 
boosted, while there is no such event in the pure noise counts, 
as shown by the dotted curve.
There are two main reasons for the enhancement
(see \S \ref{sec:lens-noisy} and 
Appendix \ref{sec:correction} for further details). 
First, addition of the noise leads to modifications of the peak height
as well as of the peak position. 
Second, the noise generates spurious peaks in the noisy maps. 
In particular, the latter effect 
is likely to show up in a biased way depending on the peak height. 
Since the convergence field around a given high peak has 
greater amplitude than the average field, adding the
noise onto such a region is more likely to induce spurious high 
peaks around the real peak.
We develop a correction scheme to the model peak counts in
Appendix \ref{sec:correction}.
The corrected counts are in reasonable agreement with the simulation
result for high peaks $\nu>5$. 

\section{Results on the correspondence between halos and peaks} 
\label{sec:result-2}

In this section, we examine correspondences between 
massive halos and peaks in the weak lensing convergence maps, using 
the results of our ray-tracing experiments and the halo catalogs created 
from the $N$-body simulation outputs. 
In doing so, we follow a realistic procedure which is to be used 
in weak lensing surveys: we focus on high peaks above a certain
threshold in the lensing map as strong candidates for halos. 
The usefulness of the weak lensing halo search can be quantified 
in terms of {\it completeness} and  {\it efficiency}.
As will be shown later in detail, choosing a threshold 
$\nu_{th}\simeq 4-5$ gives an optimal balance between the efficiency and 
completeness. 
We will also carefully examine if there is a correlation between 
the detectability of halos and their shape, because it
could produce
a bias in halo catalogs constructed from the weak lensing method.

In order to make realistic weak lensing survey catalogs, 
we impose a criterion on the field selection of simulations
so that no nearby massive halo is included in the
target fields. Specifically, we discard simulated fields 
if there is a halo (or possibly more) with 
$M_{\rm halo}>10^{14}h^{-1}M_{\odot}$ at $z<0.1$.
Similar criterion, such as no known nearby X-ray luminous cluster of 
galaxies or no nearby optically identified massive cluster, 
is usually imposed on the field selection of an actual observation,
because such a nearby massive cluster occupies a large portion of 
a survey field and makes a distant halo search difficult and inefficient.
31 out of the 50 simulated fields pass this criterion.
In addition, we discard one field in which there is a massive halo 
with $\nunfw>5$ having $flag=2$ (see \S \ref{halocatalog} for a description
of $flag$)\footnote{Including this field in our analysis below 
does not change the conclusions, but we decided 
to discard it for having conclusive results.}.
We analyze the remaining 30 fields.
We note that we do not use data within 5 arcmin of the edge of each
simulated field because of
the incomplete smoothing due to the lack of data at the outer region.
The total field size is then 106 square degrees.

\subsection{Matching halos with peaks in lensing mass maps} 
\label{sec:matching}

\begin{table}
\caption{Classification of matching relations between halos and peaks.
The double-lined arrows denote the primary match, while the single-lined 
arrows denote the secondary match. The last two columns summarize 
the number of halos/peaks in each class, the third (fourth) column is for 
matching between the halo catalog and the peak list from the noise-free 
(noisy) $\kappa$ map. Note that halos with $\nunfw>3$ (856 halos in total)
and peaks with $\nulens>3$ (669 peaks) and $\nunoisy>3$ 
(2417 peaks) are considered.}
\label{table:class}
\begin{tabular}{lccc} 
\hline
class & matching relation & \multicolumn{2}{c}{number of objects}  \\
{} & {} & $\nunfw$-$\nulens$ & $\nunfw$-$\nunoisy$ \\
\hline
i & halo${}_{\Longleftarrow}^{\Longrightarrow}$ peak & 475 & 624\\
ii & halo with no paired peak & 367 & 222 \\
iii & peak with no paired halo &  177 & 1637\\
iv  & halo${}_{\longleftarrow}^{\Longrightarrow}$ peak & 14 & 10\\
v  & halo${}_{\Longleftarrow}^{\longrightarrow}$ peak & 17 & 156\\
\hline
\end{tabular}
\end{table}

We shall now examine relations between the peak height in weak 
lensing maps and the properties of halos through the {\it expected} 
peak height for the NFW halos ($\nunfw$, see \S \ref{sec:NFWlens}).
We restrict our discussion to halos with
$\nunfw>3$ only. The halo catalogs contain 856 such halos 
in total. On the other hand, the ray-tracing simulations directly 
enable us to find peaks in the $\kappa$ maps.
We consider two kinds of the $\kappa$ maps; those with and without the
noise due to the intrinsic ellipticities.  
The peaks in the two maps 
are identified in the manner described in \S \ref{sec:result-1}.
The noise-free maps contain 669 peaks with $\nulens>3$, while the noisy
maps contain
2417 peaks with $\nunoisy>3$. Thus, the noise significantly enhances the
peak counts.  
It is also worth noting that
the 669 peaks in the noise-free maps is significantly smaller 
than the {\it expected} number of peaks from the halo catalogs.
We will carefully investigate this discrepancy in the subsequent sections. 

We consider halos and peaks with $\nu>3$ only
in order to avoid mismatches between peaks and physically unrelated halos.
Because of large number density of low mass halos, 
including such halos rapidly increases the 
chance of mis-matching.
This is especially the case when one attempts to match 
peaks in the noisy $\kappa$ maps with halos because 
false peaks due to the noise
can be accidentally matched with physically unrelated halos.
The criterion $\nu>3$ is chosen based on the fact that most of such 
high peaks are expected to be caused by real halos, even when the 
lensing map is contaminated by the noise 
(see Figure \ref{fig:peakcounts}).
Also the criterion value is reasonably lower than the observational 
threshold value which is generally set by $\nu \ga 4$.
This guarantees most halos of interest to us are included in 
the final halo catalog.

We carry out the peak-halo matching 
in two directions: whether a peak has a corresponding
halo in the halo catalog, and whether a halo has a counter-peak in the
simulated $\kappa$ map.
Note that the peak position does not necessarily agree with 
the halo's center-of-mass position exactly and 
this is more likely the case for halos that have substructures or are
in the merger process. For this reason,  we search for a 
matched pair candidate within a radius of 12 pixels (2.88 arcmin) 
from the peak position or from the halo center.
This maximum angular separation 2.88 arcmin is chosen so that it is larger 
than the smoothing radius of $\theta_G=1$ arcmin, while it is still smaller 
than the angular virial radius of a massive halo at $0.1 \la z \la 0.7$. 
If there are more than one pair candidate within the radius, 
we take the closest one as the primary candidate and consider the others
as the secondary candidate.

We classify the peak-halo matching into the following 
five classes:
Class-(i) is that the both directions agree; both a halo and 
a peak are their primary pair candidate. 
Class-(ii) is defined as halos that have no paired peak within a radius of 
12 pixels from the halo center.
We will call such halos as {\it ``missing halos''}.
Class-(iii) is defined as peaks with no paired halo within a radius of 
12 pixels from the peak position.
We will often refer to such peaks as {\it ``false peaks''}.
Class-(iv) denotes a halo that has a matched peak candidate, but 
is not the primary matched halo of that peak.
Class-(v), which is the opposite to the class-(iv), is that 
a peak has a matched halo candidate but is not the primary matched peak 
of that halo. 
The resulting numbers for the noise-free and noisy maps are 
summarized in Table \ref{table:class}.

Typical properties of halos or peaks in the class-(ii--v) are explained
as follows.
The class-(ii) halos are those which have a peak height lower than
the criterion value ($\nu=3$) though the expected peak height, $\nunfw$,
exceeds it.
This is mainly due to departures from the NFW profile
or/and the projection effect.
In the class-(iii) peaks there are two kinds:
The first is the peaks which have corresponding halos but have the
expected peak heights, $\nunfw$, lower than the criterion value, though
the measured peak heights in a $\kappa$ map exceed it.
This is again due to departures from the NFW profile or/and the 
projection effect.
The other is the peaks which do not have corresponding halos but
are generated by the noise.
The former accounts for most of the class-(iii) peaks in noise-free 
$\kappa$ maps, while the latter accounts for a large fraction of those
in noisy $\kappa$ maps.
The class-(iv) halos are due to the projection effect; 
if two or more halos are located at different redshifts
along almost the same line-of-sight, after smoothing of the $\kappa$ map
only one peak appears, because the smoothing merges those into one structure. 
Finally, in the class-(v) peaks there are two kinds:
The first is the peaks due to substructures within a halo. 
This accounts for most of the class-(v) peaks in noise-free $\kappa$ maps. 
The other is false peaks due to the noise generated around the primary peak 
due the massive halo.
Because the convergence field around the high peak is biased and  
thus larger than the average field, adding the noise into such
region is more likely to generate false peaks with a relatively high 
peak height. 
This accounts for most of the class-(v) peaks in noisy $\kappa$ maps.

In the rest of this section, we consider the class-(i--iii) objects only,
and do not deal with the class-(iv) and (v) objects,
because the class-(iv) and (v) objects are the secondary
matched pairs associated with the primary matched pair in the 
vicinity.
This roughly mimics the actual observation procedure
in which the secondary objects are investigated in detail in course of a
follow-up observation of the primary matched object.

\subsection{Halos versus peaks in noise-free $\kappa$ maps} 
\label{sec:halo-lens}

\begin{figure}
\begin{center}
\begin{minipage}{8.4cm}
\epsfxsize=8.4cm
\epsffile{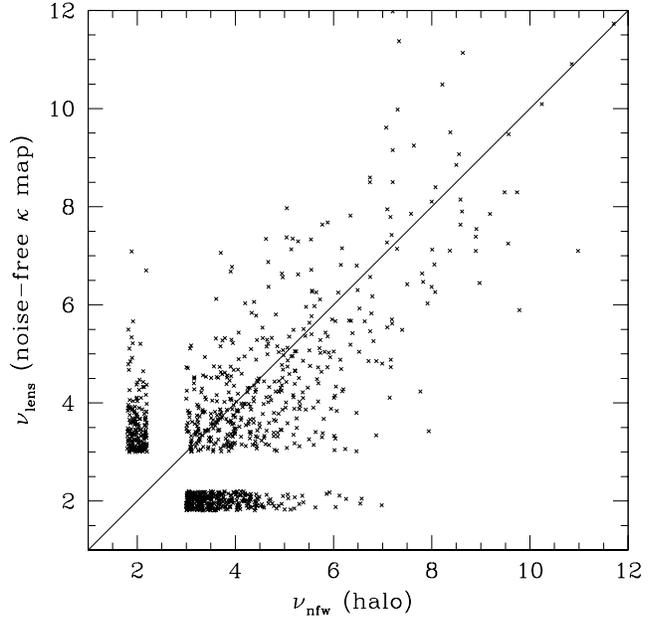}
\end{minipage}
\end{center}
\caption{The correspondences between halos and 
peaks identified in the noise-free $\kappa$ maps in $\nunfw$-$\nulens$ plane.
The class-(i-iii) objects are shown.
The class-(ii) objects are placed within a narrow track around 
$\nulens\sim 2$, regardless of their true value, and the class-(iii) 
objects are placed at $\nunfw\sim 2$ as well.}
\label{fig:halo-noisefreepeak}
\end{figure}

Let us first look into the correspondence between high peaks identified
in the noise-free $\kappa$ maps ($\nulens$) and properties of halos 
through the {\it expected} peak height ($\nunfw$).
The main purpose of this subsection is to examine the amplitude of the 
irreducible scatter in the relation induced by both the 
individuality of halo's mass distribution and the projection effect.

\begin{figure}
\begin{center}
\begin{minipage}{8.4cm}
\epsfxsize=8.4cm 
\epsffile{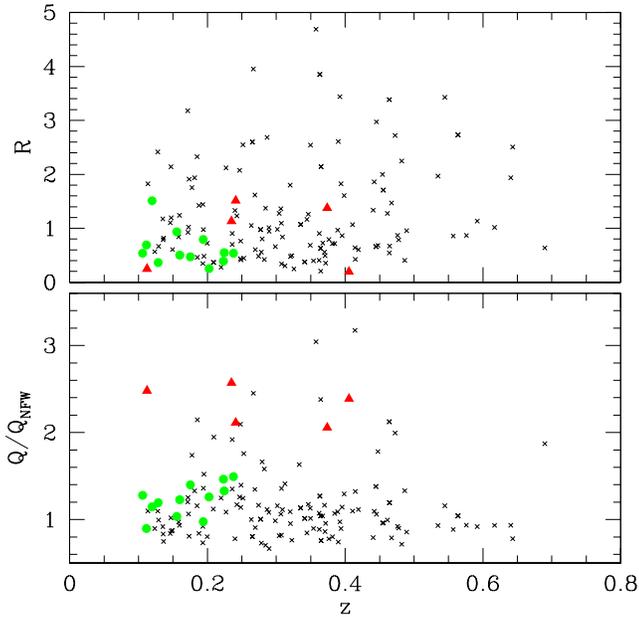}
\end{minipage}
\end{center}
\caption{The scatter plot for the distribution of the halos 
in $z$--$Q/Q_{\rm NFW}$ plane (lower panel) and in $z$--$R$ plane 
(upper panel).
The filled circle and triangle symbols are for 
the class-(ii) halos with $\nunfw>5$.
For comparison, the class-(i) halos with $5<\nunfw<7.5$ are also 
plotted by the cross symbols. 
It is apparent from the lower panel that the class-(ii) halos can be
separated into two subclasses:
those with $Q/Q_{\rm NFW}>2$ are plotted by the filled triangles,
while the rests are plotted by the filled circles.}
\label{fig:z-q-r-class2}
\end{figure}

Figure \ref{fig:halo-noisefreepeak} 
shows the distribution of the class-(i--iii) objects 
in the $\nunfw$-$\nulens$ plane.
Since the class-(ii) halos do not have 
a matched peak with height $\nulens>3$ in the $\kappa$ map 
(although the halos have a high expected peak height of $\nunfw>3$), 
such halos are placed in a narrow locus around $\nulens=2$ for
illustrative purpose.
For the similar reason,  the class-(iii) peaks are placed around 
$\nunfw = 2$.
The distribution of the class-(i) halos shows a large scatter around
the ideal relation $\nulens=\nunfw$ (shown by the solid line in the 
upper panel) over the range of $\nu$ we consider. 
It is worth noting that the distribution for a fixed 
$\nunfw$ value is slightly biased toward lower $\nulens$ 
values. The mean and RMS of the differences, 
$\nulens-\nunfw$, are $-0.24$ and 1.2, respectively.
Since no noise is added to the $\kappa$ map in this case, 
the scatter and bias must be ascribed to the two effects;
(1) the mass distribution within a halo is not spherical nor
completely universal, causing deviations of the peak heights from
the expected values computed from our simple model, and (2)
a chance projection of structures along the same line-of-sight alters 
the lensing signal due to the halo. 
For the former, although we have so far used the average relation for the 
halo concentration, given by eq.~(\ref{cnfw}), it was shown in Jing (2000) 
that the halo
concentrations among halos with a given mass obey the log-normal
distribution with the variance $\sigma(\ln \cnfw)\approx 0.2$.
This roughly corresponds to a 20\% RMS scatter in the $\cnfw$ 
relation. 
Given the approximate relation $\kappa \propto \cnfw^{1-1.5}$ for the
peak height due to a massive halo (see \S \ref{sec:models}), the RMS 
scatter in the $\cnfw$ distribution translates into $20-30\%$ RMS scatters 
in the lensing signal $\nunfw$ and therefore equivalently
in the $\nulens$-$\nunfw$ relation.  
For example, for halos with $\nunfw=5$, 
the RMS of $\sigma_\nu \sim 1$ is caused.
Second, the lensing projection due to physically unrelated structures 
along the same line-of-sight sight of
a halo also contributes to the scatter in $\nulens$. 
Following Hoekstra (2001), this scatter can be estimated from
the cosmic weak lensing variance under the assumption that  
the halo and the large-scale structures are uncorrelated.
For the Gaussian smoothing scale $\theta_G=1'$, the amplitude of 
the (linear) convergence variance is  $\sigma_\kappa (1')=0.01$ for the
$\Lambda$CDM model,
corresponding to the RMS deviation of $\sigma_\nu=0.5$ because of the 
noise $\sigma_{\rm noise}(1')=0.02$ (see eq.~[\ref{sigma-noise}]).  
Hence, these two effects can basically explain the amplitude of 
the scatter in the 
$\nulens$-$\nunfw$ relation in Figure \ref{fig:halo-noisefreepeak}.

We also argue that the systematic bias in the $\nulens-\nunfw$ relation 
can be explained by the following two effects. 
The first is the fact that more than half of the halos have $R<1$ 
as discussed in \S \ref{halocatalog} and Figure \ref{fig:PDF-R}.
Since $R$ positively correlates with $\nulens/\nunfw$ 
(see Figure \ref{fig:r-nu-class1} in Appendix \ref{sec:haloshape}),
it causes the bias.
The second is the projection with a void region in the same line-of-sight 
which reduces the lensing signal (see Metzler et al.~2001; 
White et al.~2002; Padmanabhan et al 2003 for detailed study on 
the projection effect).

\begin{figure}
\begin{center}
\begin{minipage}{8.4cm}
\epsfxsize=8.4cm 
\epsffile{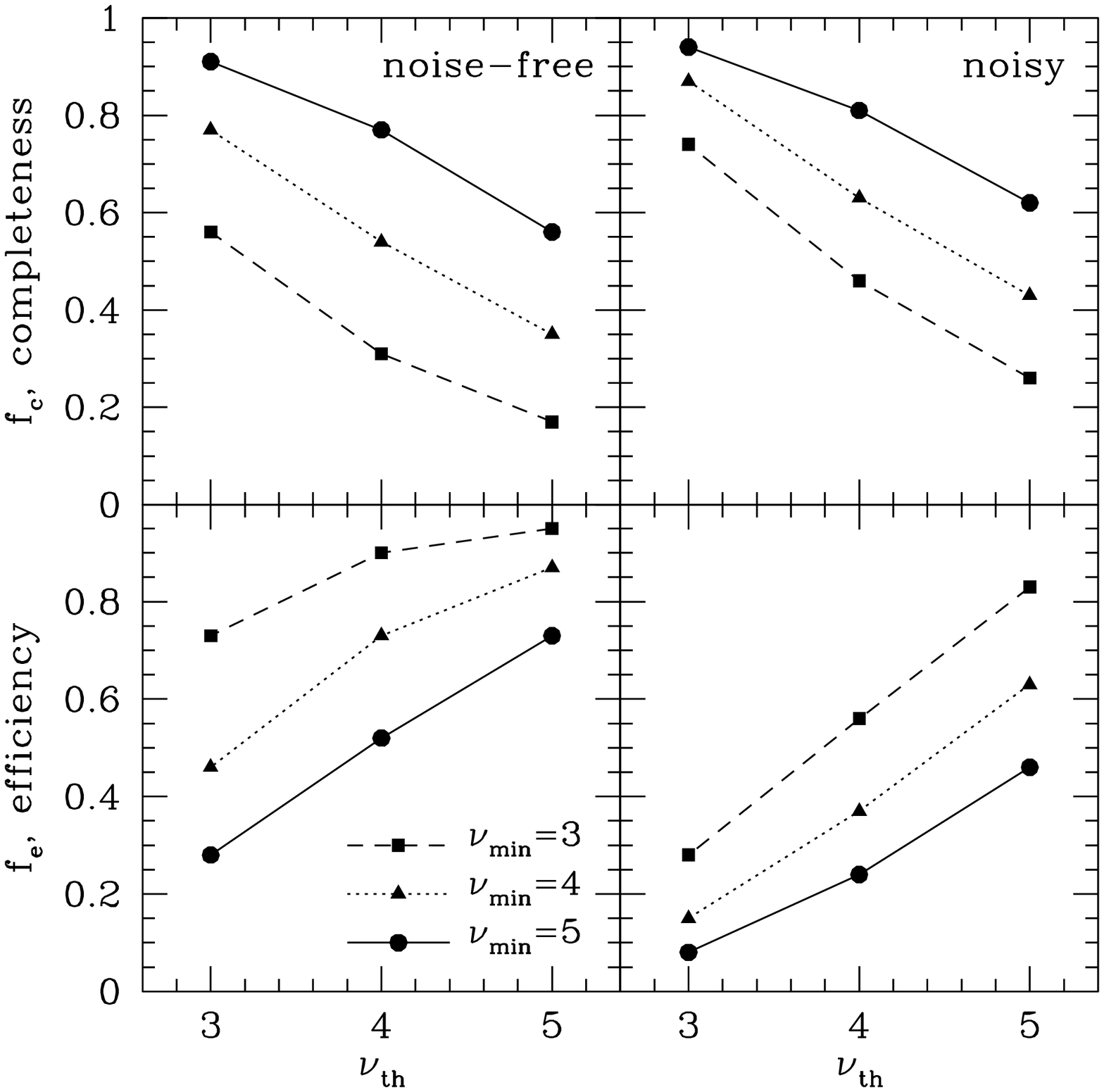}
\end{minipage}
\end{center}
\caption{{\it Upper Panels}: The completeness 
$f_c(\nu>\nuth|\nunfw>\numin)$ defined by eq.~(\ref{completeness}),
as a function of the threshold peak height $\nuth$. 
Three cases for the minimum expected peak value, $\numin=3$, 4 and 5
 are shown by the filled squares, triangles and circles, respectively.
{\it Lower Panels}: The efficiency
$f_e(\nunfw>\numin|\nu>\nuth)$ defined by eq.~(\ref{efficiency}),
as a function of the threshold peak height $\nuth$.
The left panels are for the noise-free case ($\nu=\nulens$), while 
the right 
panels are for the noisy case ($\nu=\nunoisy$).}
\label{fig:f}
\end{figure}

\begin{figure*}
\begin{center}
\begin{minipage}{15cm}
\epsfxsize=15cm 
\epsffile{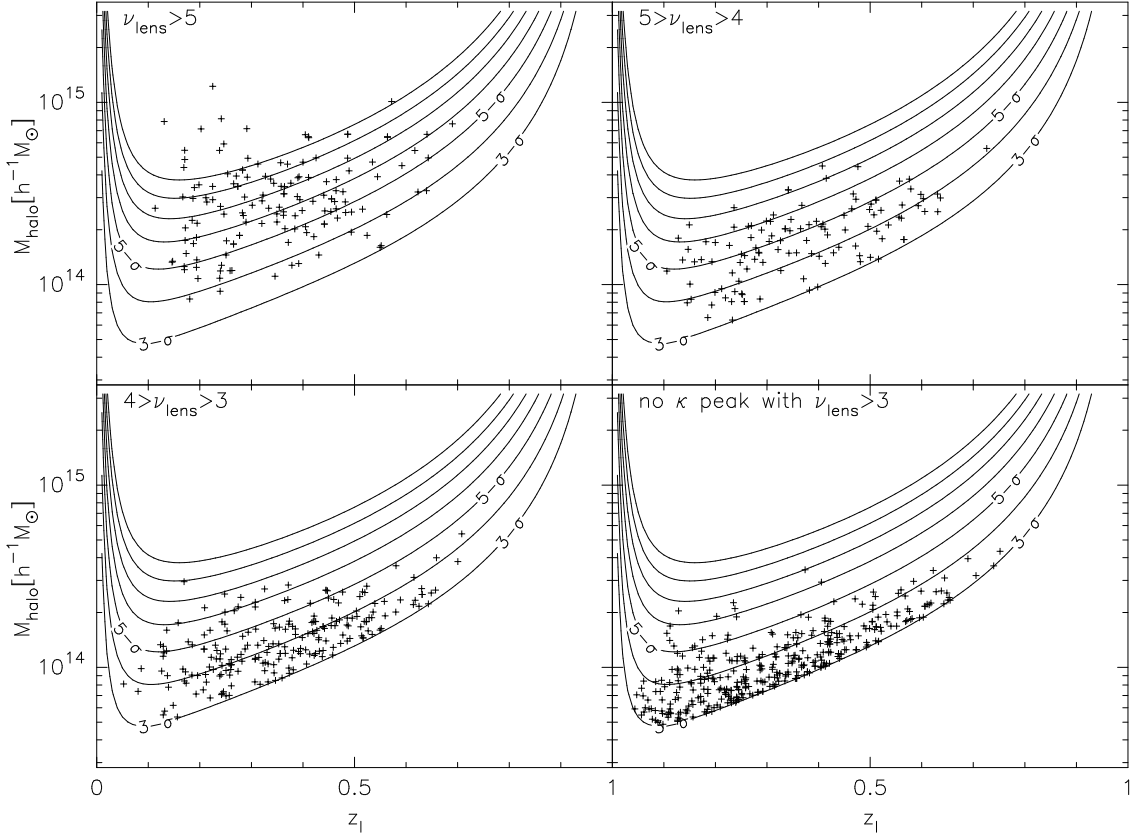}
\end{minipage}
\end{center}
\caption{The scatter plot for the halos in the redshift-halo mass plane.
The contours show the {\it expected} peak height $\nunfw$ from 3 to 9 at an
interval of $\Delta\nunfw=1$.
The class-(i) halos matched with peaks identified in {\it noise-free} $\kappa$
maps are plotted in the upper-left ($\nulens>5$), 
upper-right ($5>\nulens>4$) and lower-left panels ($4>\nulens>3$). 
The lower-right panel shows
the scatter plot for the class-(ii) ``{\it missing halos}''.}
\label{fig:scatter-noisefree}
\end{figure*}

Let us turn to the class-(ii) halos in Figure
\ref{fig:halo-noisefreepeak}.
Among the 367 class-(ii) halos, 
there are 17 halos that are supposed to produce peaks with $\nu>5$ in
the $\kappa$ maps if the halos have the NFW profile 
(i.e., $\nunfw>5$ in our notation).
Why are these halos missing?
Figure \ref{fig:z-q-r-class2} explains
this in terms of deviation of the mass
distribution within a halo from the spherically symmetric NFW profile.
We again use the parameters $Q$ 
and $R$ to quantify the mass distribution within a halo 
(see equations (\ref{Q}) and (\ref{R}), respectively):
$Q/Q_{\rm NFW}<1$ $(>1)$ characterizes a halo that has centrally 
more (less) concentrated
mass distribution than expected from the NFW profile, 
while $R>1$ $(<1)$ describes that the mass distribution is
elongated along (perpendicular) to the line-of-sight ($R=1$ for the 
spherical symmetric halo). 
Therefore, a halo
with $Q/Q_{\rm NFW}>1$ and/or $R<1$ likely leads to
a smaller lensing signal than expected
from the NFW profile (see Appendix \ref{sec:haloshape} for more detailed 
analysis of the relations between the halo shape and the amplitude 
of the lensing signal).  
The upper panel of Figure \ref{fig:z-q-r-class2} shows
the distribution of the class-(ii) halos
in $z$--$Q/Q_{\rm NFW}$ plane, while the lower panel shows the distribution
in $z$--$R$ plane. For comparison, the figure also shows the 
distributions for 
the class-(i) halos with $5<\nunfw<7.5$ by the cross symbols.
{}From the lower panel, it is clear that the class-(ii) 
halos can be divided into 
two distinct subgroups; those which have $Q/Q_{\rm NFW}>2$ 
(plotted by the filled 
triangles) or $Q/Q_{\rm NFW}<2$ (plotted by the filled circles).
We indeed find that halos with $Q/Q_{\rm NFW}>2$ show clear signatures
of substructures or on-going merging, resulting in the extended mass
distribution 
(the $\kappa$ map of such halos is shown in Figure 
\ref{fig:missing-halo-1}, for example).
The remaining objects with $Q/Q_{\rm NFW}<2$ tend to be located at a
low redshift ($z<0.25$), where the lensing efficiency is relatively
small for the source redshift $z_s=1$ we consider, and have a small $R$, 
mostly $R<1$.
The examples are shown in Figures \ref{fig:missing-halo-2} and 
\ref{fig:missing-halo-3}.
These halos typically 
have prominent substructures and the whole mass
distribution within the halo
is indeed elongated along the perpendicular direction to the line-of-sight.
Except for such irregular systems, 
almost all massive halos with $\nunfw>5$ generate a high peak with 
$\nulens>3$.
It is important to note that the distribution of the class-(ii) halos in 
Figure \ref{fig:z-q-r-class2} is not clearly distinguished from that of the 
class-(i) halos, reflecting the fact that the lensing
signal is also affected by the projection effect.

Turn next to the class-(iii) objects in Figure
\ref{fig:halo-noisefreepeak} which are the peaks with
$\nulens\ge3$ in the lensing map, but do not have the corresponding
halos in the halo catalog (contains only halos with $\nunfw>3$).
Among the class-(iii) objects
there are 8 events with $\nulens>5$ which we call ``false peaks''.
We carefully looked into each case in the $\kappa$ map and searched
for a possible related halo in the halo catalog. We found 
that the 5 events 
are associated with substructure of massive halos or the secondary 
local peaks of merger systems (note that the peaks are separated from 
the primary peak of the halo center by more than our searching distance,
$12 {\rm pixels} = 2.88$ arcmin).
Figure \ref{fig:local-peak} shows the $\kappa $ map of one of such cases. 
The remaining 3 peaks are due to a chance projection of two moderately 
massive halos at different redshifts 
(Figure \ref{fig:chance-projection} shows the one example).
Therefore, except for the three chance projection systems,
the {\it false peaks} with $\nulens>5$ are all associated with a 
real massive halo as the secondary peak.

In order to quantify the ability of weak lensing survey 
to locate massive halos, we consider the {\it completeness} 
and {\it efficiency}.
We define the {\it completeness} by the fraction of halos, 
which are detected as high peaks above a given threshold $\nuth$ in the 
$\kappa$ map, relative to all the halos with the expected lensing signal, 
$\nunfw$, greater than a given minimum value $\numin$:
\begin{equation}
\label{completeness}
f_c(\nulens>\nuth|\nunfw>\numin)
={{N_{\rm i}(\nulens>\nuth|\nunfw>\numin)} 
\over {N_{\rm i+ii}(\nunfw>\numin)}},
\end{equation}
where $N_{\rm i+ii}(\nunfw>\numin)$ denotes the number of halos with
$\nunfw>\numin$ which, in our classification, corresponds to the number of 
the class-(i) and (ii) such halos, and 
$N_{\rm i}(\nulens>\nuth|\nunfw>\numin)$ denotes the halos having
$\nunfw>\numin$ matched with the peaks with $\nulens>\nuth$ (thus 
these are the class-(i) objects).
Similarly, we define the {\it efficiency}
by the fraction of peaks, which are associated with halos with the expected
peak height exceeding a given minimum value $\numin$, relative to all the 
peaks with height greater than a given threshold $\nuth$.
\begin{equation}
\label{efficiency}
f_e(\nunfw>\numin|\nulens>\nuth)
={{N_{\rm i}(\nunfw>\numin|\nulens>\nuth)} 
\over {N_{\rm i+iii}(\nulens>\nuth)}},
\end{equation}
where $N_{\rm i+iii}(\nulens>\nuth)$ denotes the number of peaks with
$\nulens>\nuth$ which corresponds to the number of 
the class-(i) and (iii) such peaks.

The results are plotted in the left panels of Figure \ref{fig:f} as a 
function of $\nuth$.
The upper panels show the completeness ($f_c$) 
and the lower panels show the efficiency ($f_e$). 
As expected, these two estimates behave inversely,
that is, when $f_e$ goes up, $f_c$ goes down.
Therefore the optimal choice of the $\nuth$ value should be 
determined by balance between the two estimates,
which we discuss in the latter section with taking into 
account effects of the noise.

The plot for the completeness shows that only about 
half of halos with $\nunfw>5$ produce peaks with $\nulens>5$ 
in the noise-free $\kappa$ maps, and that 77 percent of the halos 
have $\nulens>4$.
The completeness becomes worse if $\nuth \ga \numin$, as expected. 
It is also shown that a good completeness (say $f_c \ga 0.8$) is 
attainable if one sets $\nuth \la\numin-1$.
This can be understood from the fact that the RMS scatter caused by 
the individuality of halos and the projection effect is 
$\sigma_\nu\sim 1.2$.
Note that some fraction of the halos with $\nulens$ much smaller than
the expected peak heights $\nunfw$ are irregular systems or 
un-relaxed systems, 
and therefore the completeness is likely to be improved for samples of
regular, well relaxed halos.
It should be also noted that the noise increases 
the completeness (see \S \ref{sec:fof-noisy} and the upper right panel).

Let us turn to the efficiency.
Since no noise is added to the $\kappa$ maps, the efficiency given here
should be regarded as an estimate for an ideal lensing survey. 
Clearly, the efficiency is very high.
If one sets the detection threshold to be $\nuth=5$, 73 percent
of all the peaks correspond to real halos with $\nunfw>5$,
and 95 percent for $\nunfw>3$.
Even for $\nuth=4$ the efficiency is still high,
90\% for $\nunfw>3$.
Therefore, in the absence of the noise, the contamination rate
(the fraction of false peaks) is very small for detection threshold 
$\nuth>4$.

\begin{figure}
\begin{center}
\begin{minipage}{8.4cm}
\epsfxsize=8.4cm 
\epsffile{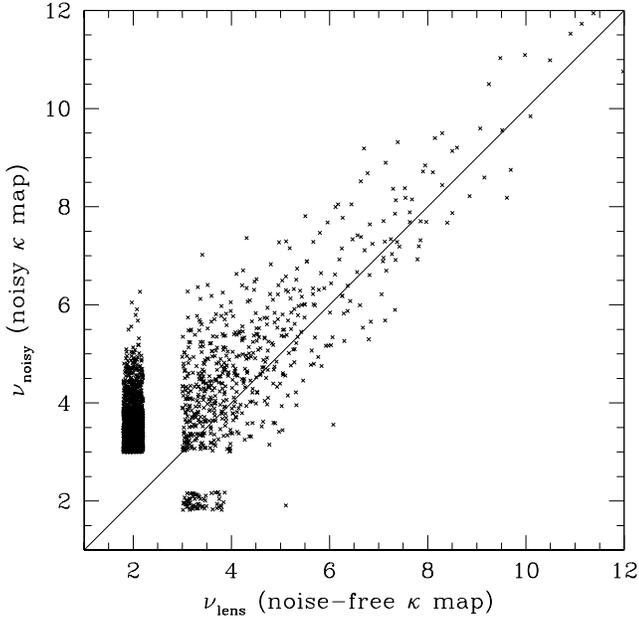}
\end{minipage}
\end{center}
\caption{The correspondences between the peak heights
identified in the noise-free and noisy $\kappa$ maps.
The peaks, which have $\nulens>3$ but are missed in the noisy
$\kappa$ map or have $\nunoisy<3$, are placed within a narrow locus at 
$\nunoisy\sim2$, regardless of their true $\nunoisy$ value.
Similarly, the peaks, which have $\nunoisy>3$ but no paired peak with 
$\nulens>3$, are placed at $\nulens\sim2$,}
\label{fig:noisefree-noisy}
\end{figure}

Figure \ref{fig:scatter-noisefree} shows the scatter plot of the class-(i) 
and -(ii) halos in the redshift--halo mass plane.
The contours show tracks of the expected 
peak height $\nunfw$ from 3 to 9 at an
interval of $\Delta\nunfw=1$.
The class-(i) halos matched with peaks in {\it noise-free} $\kappa$
maps are plotted in the upper-left ($\nulens>5$), 
upper-right ($5>\nulens>4$) and lower-left panels
($4>\nulens>3$), respectively. 
The lower-right panel shows 
the scatter plot for the class-(ii) halos, namely ``{\it missing halos}''.
Importantly, 
halos within a small range of the peak height are spread
over a wide range of masses  
due to the projection effect and the individuality of the halo mass 
distribution.

\subsection{Effects of the noise on the lensing peaks}
\label{sec:lens-noisy}

Before examining correspondences between peaks identified in the noisy 
$\kappa$ maps and halos in the halo catalog, 
we quantitatively examine 
how the noise alters the peaks due to halos and generates 
false peaks in the lensing map. 

Matching between the noise-free peaks and the noisy peaks was done in the 
same manner described in \S \ref{sec:matching}.
In what follows, the peaks classified in the class-(i-iii) (its analogy to 
$\nulens$-$\nunoisy$ correspondence) are considered.
Again, we only deal with peaks with $\nu>3$.

Figure \ref{fig:noisefree-noisy}
shows the relations between $\nulens$ and $\nunoisy$ plotted
in the same manner as in Figure \ref{fig:halo-noisefreepeak}.
We emphasize that the statistical properties of 
the $\nulens$-$\nunoisy$ relation do not follow the Gaussian 
distribution, even though the Gaussian noise is added.
The reason is as follows.
The noise not only alters the peak height from massive halos
but also changes the peak position.
For example, if a large positive noise is accidentally added on 
a neighboring pixel around the original peak position, the
pixel could be identified as a peak that is even higher than the
original peak.
Therefore, the distribution of $\nunoisy$'s is not simply expressed as
the convolution of that of $\nulens$'s with the Gaussian noise, 
but the $\nulens$-$\nunoisy$ relation becomes asymmetric
and is biased toward large $\nunoisy$ values.
This explains the asymmetric $\nulens$-$\nunoisy$ relation clearly 
shown in Figure \ref{fig:noisefree-noisy}.
The mean and RMS of the differences, $\nunoisy-\nulens$, 
among the class-(i) objects with $\nulens>4$ (this restriction is 
imposed in order to avoid an incomplete 
sampling of low $\nunoisy$ pairs) are 0.31 and 0.90, respectively.
Therefore the peak heights of massive halos identified in a 
noisy $\kappa$ map are, on average, boosted by the noise.
This bias partly accounts for the excess in the peak number counts 
over the simple theoretical prediction, eq.~(\ref{defcounts}), as
discussed in \S \ref{sec:peakcounts}.

\begin{figure}
\begin{center}
\begin{minipage}{8.4cm}
\epsfxsize=8.4cm 
\epsffile{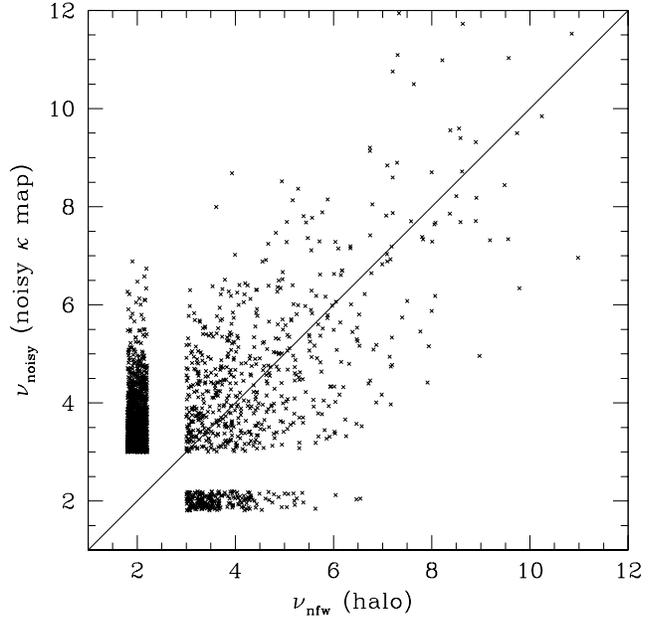}
\end{minipage}
\end{center}
\caption{Same as Figure \ref{fig:halo-noisefreepeak} but for the 
correspondence between halos and peaks identified in the noisy
$\kappa$ maps.}
\label{fig:halo-noisypeak}
\end{figure}

\begin{figure*}
\begin{center}
\begin{minipage}{15cm}
\epsfxsize=15cm 
\epsffile{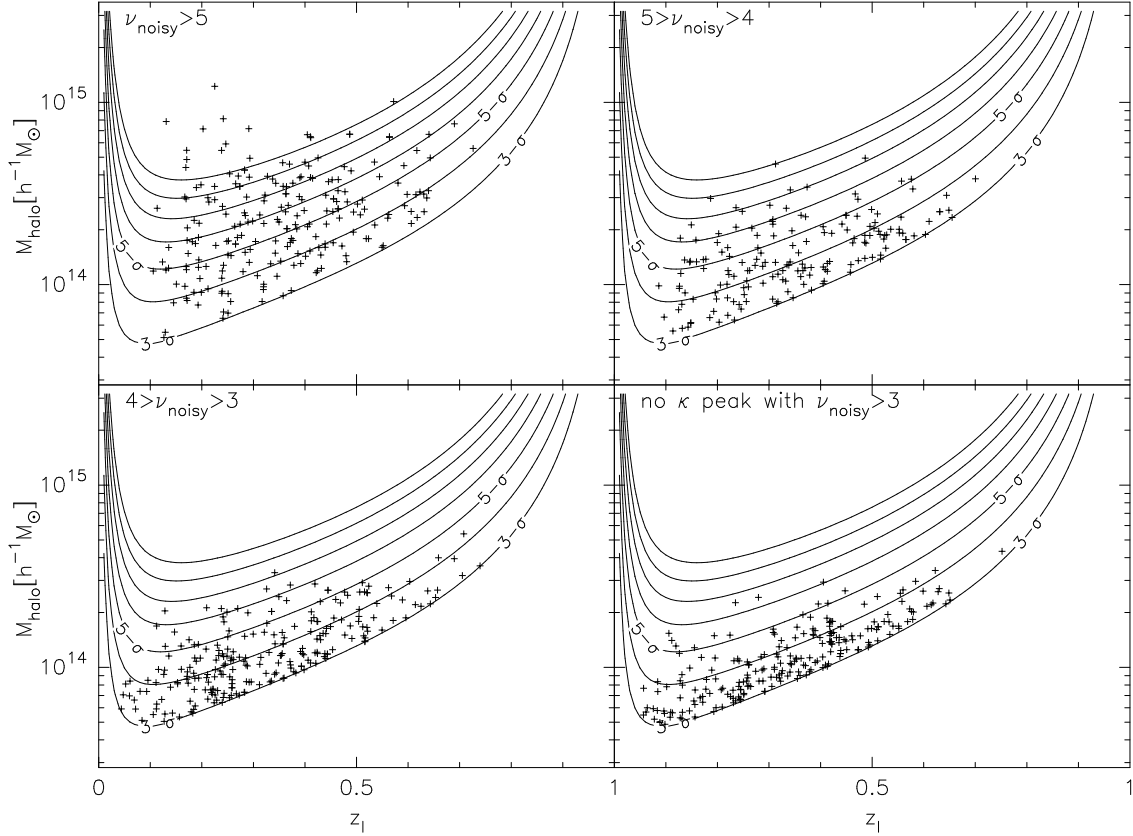}
\end{minipage}
\end{center}
\caption{Same as Figure \ref{fig:scatter-noisefree} but for halos 
matched (not matched, for lower-right panel) with peaks identified in 
{\it noisy} $\kappa$ maps.}
\label{fig:scatter-noisy}
\end{figure*}

The class-(ii) objects here correspond to 
the {\it missing peaks}, the peaks that have $\nulens>3$ but are 
absent in the corresponding noisy $\kappa$ map or have $\nunoisy<3$.
Except for one peak with $\nulens=5.1$, the missing peaks sit in the range
of $\nulens\le 4$. These peaks are likely to be erased or their 
heights are decreased by the addition of the negative noise.
The one exception (the peak with $\nulens=5.1$ at the missing peak locus)
is a local peak associated with a substructure of a massive halo which
disappears in the noisy $\kappa$ map.
We may, therefore, safely conclude that more than 80$\%$ of the peaks
with $\nulens\ge 3$ in the noise-free map remain in the noisy $\kappa$ map.

Next, we consider the {\it false peaks}, 
the peaks that are identified in the noisy 
$\kappa$ map with $\nunoisy>3$ but does not correspond to peaks 
in the noise-free $\kappa$ map or have $\nulens<3$, 
which are placed around $\nulens\sim2$ in Figure \ref{fig:noisefree-noisy}.
A large fraction of the false peaks, 
especially for $\nunoisy<4$, are the false peaks 
generated by the noise (see \S \ref{sec:result-1} for details), and the
small fraction (especially those with relatively high $\nunoisy$) is 
associated with real peaks on which a large noise is accidentally added.
Actually, there are 14 objects with $\nunoisy>5$ in the {\it false peaks}
locus. 
All of those have a corresponding peak in the noise-free $\kappa$ map with
low peak height $1.5<\nulens<2.9$ on which a relatively large noise 
of $2\sim 3.5\sigma$ is accidentally added.
Therefore, all peaks with $\nunoisy>5$ identified in the noisy $\kappa$ map 
are associated with real peaks in the noise-free $\kappa$ map.
On the other hand, most of 
the class-(iii) peaks with lower heights are 
false peaks generated by the noise. 
In fact, one can find 
the rapid increasing number of the false peaks toward low 
peak heights, especially for $\nunoisy<4$, as shown in \S \ref{sec:result-1}.
It immediately follows from this that in order to reduce the  
contamination rate, a high threshold value, $\nuth \ga 4$ at least, 
is required.

\subsection{Halos versus peaks in noisy $\kappa$ maps}
\label{sec:fof-noisy}

\begin{figure}
\begin{center}
\begin{minipage}{8.4cm}
\epsfxsize=8.4cm 
\epsffile{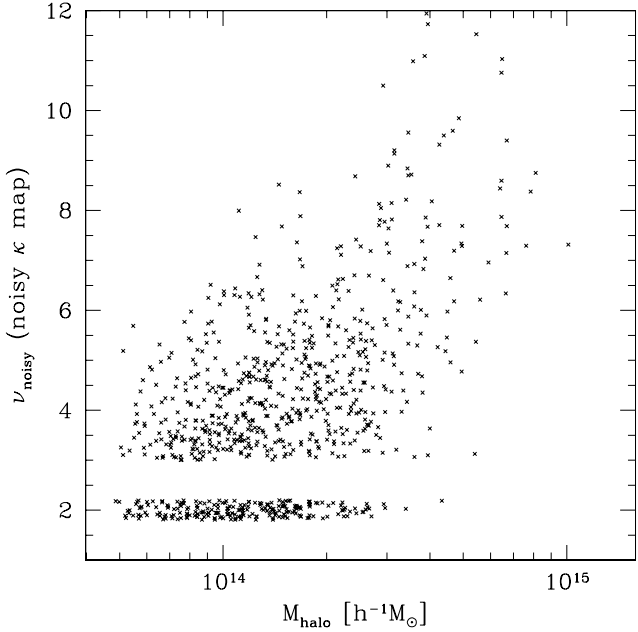}
\end{minipage}
\end{center}
\caption{Same as Figure \ref{fig:halo-noisypeak} but the halo mass
is used instead of the {\it expected} peak height of $\nunfw$. 
Note that only halos with $\nunfw>3$ are considered.}
\label{fig:halomass-noisypeak}
\end{figure}

In this subsection, we
examine correspondences between halos in the halo
catalog and peaks in the $\kappa$ map including the noise due to
the intrinsic ellipticities.
Figure \ref{fig:halo-noisypeak} shows the distribution of the class-(i-iii) 
objects in $\nunfw$-$\nunoisy$ plane plotted in
the same manner as in Figure \ref{fig:halo-noisefreepeak}.
The comparison with Figure \ref{fig:halo-noisefreepeak} reveals
that the noise not only increases the scatter in the 
$\nunoisy$-$\nunfw$ relation but also causes the systematic bias in that
the peak heights in the noisy map are statistically boosted, 
as seen in the previous subsection.
Interestingly, this bias leads the {\it completeness} for the weak
lensing halo search to be improved. The upper-right and lower-right
panels in 
Figure \ref{fig:f} show the completeness 
and efficiency,  defined in the same manner
as eqs.~(\ref{completeness}) and (\ref{efficiency}), for the noisy maps.
About 81 (63) percent of massive halos with $\nunfw>5$ ($>4$) are 
identified as peaks with $\nu_{\rm noisy}\ge 4$ in the noisy maps.
This is one of the most important results of this paper, indicating that 
a high completeness of the weak lensing halo search could 
be attained in a realistic data.

There are 13 {\it missing} halos that have $\nunfw>5$ but $\nunoisy<3$
or have no corresponding peak in the noisy $\kappa$ map.
We look into each case in the $\kappa$ map and find that 11 halos out of 
13 are 
irregular systems such as un-relaxed systems having substructures
or on-going merger.
Therefore, the {\it completeness} of the weak lensing halo search 
for regular, well relaxed systems would be higher than the estimate 
plotted in Figure \ref{fig:f}.
The remaining 2 {\it missing} halos, which have $\nunfw=5.3$ and 5.4, are 
halos which accidentally meet a large negative noise so that their $\kappa$ 
values are below $\nunoisy=3$.
The fraction of such halos over all halos with $\nunfw>5$ is 0.8\% 
(2 out of 250 halos), which is indeed very small.
Hence, we can safely conclude that almost all halos with $\nunfw>5$ 
generate peaks with $\nunoisy>3$ even in the presence of the noise.

Let us turn to the class-(iii) objects.
Since the noise generates false peaks in the $\kappa$ map, 
the {\it efficiency} becomes 
worse than the noise-free case (see Figure \ref{fig:f}). 
The fraction of contamination by the false peaks is larger for a lower peak
height because the noise generates relatively lower peaks, mostly
$\nunoisy<4$. Therefore, if $\nuth=3$ is employed as the detection threshold, 
only 28\% of all peaks are associated with halos with $\nunfw>3$.
Note that since we only consider halos with $\nunfw>3$, some fraction of 
{\it missing peaks} could be associated with halos with $\nunfw<3$.
On the other hand, the efficiencies for higher detection thresholds 
are reasonably high: for $\nuth=5$ ($\nuth=4$), 85\% (56\%) of all peaks
with $\nunoisy>\nuth$ are associated with halos with $\nunfw>3$.

\begin{figure}
\begin{center}
\begin{minipage}{8.4cm}
\epsfxsize=8.4cm 
\epsffile{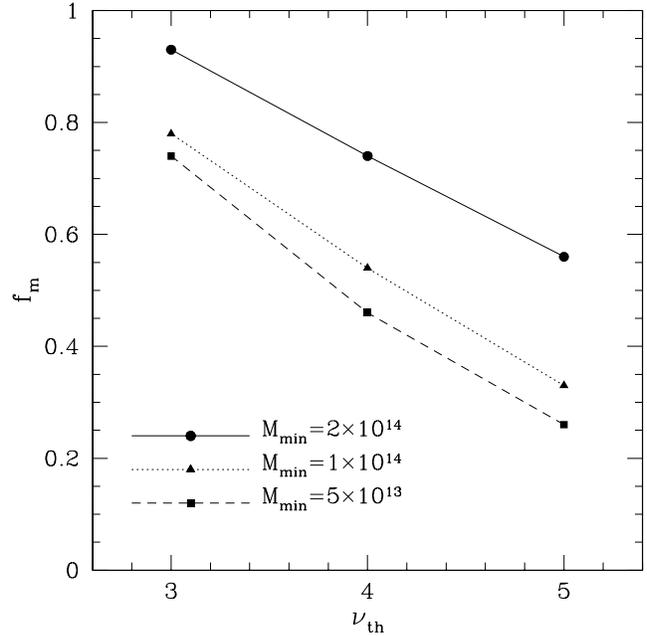}
\end{minipage}
\end{center}
\caption{The number fraction of halos identified by peaks with 
height higher than $\nuth$ relative to 
all the halos with mass 
greater than a given minimum mass, $M_{\rm min}$,
$f_m(\nunoisy>\nuth|M_{\rm halo}>M_{\rm min})$ (see 
eq.~(\ref{fm})). 
Three cases of $M_{\rm min}=5\times 10^{13}$, $1\times 10^{14}$, 
and $2\times 10^{14}h^{-1}M_\odot$ are plotted.
Note that only halos with $\nunfw>3$ are considered.}
\label{fig:fm}
\end{figure}

{}From Figure \ref{fig:f}, it is clear that
a better {\it efficiency} achieved by setting 
a higher detection threshold is 
a ``trade-off'' for a worse {\it completeness}.
We also note that the peak counts varies sensitively with the 
threshold. For example, the counts $N(\nu>4)$ is about twice 
as large as $N(\nu>5)$ (see \S 4).
Hence, the detection threshold should 
be determined such that the {\it efficiency} and 
{\it completeness} are optimized for specific survey strategies.
The results we have so far shown indicate that  
$\nunoisy\simeq 4-5$ is reasonably optimal
for currently typical weak-lensing surveys.
To be more specific, 
the results from the simulated $\kappa$ maps in the $\Lambda$CDM model 
indicate that, the mean number of halos with $\nunfw>4$
is $N_{\rm halo}(\nunfw>4)=37$ per 10 deg$^2$, 
and on average  63\% of such halos (23 halos out of 37) are detected 
with $\nunoisy>4$.
The completeness is better for higher $\nunfw$ halos: for halos with
$\nunfw>5$, the mean number is $N_{\rm halo}(\nunfw>5)=19$ per 10 
deg$^2$, where 81\% of the halos (15 halos out of 19) are detected 
with $\nunoisy>4$, and 12 halos have even high lensing signal of 
$\nunoisy>5$. 
Concerning the efficiency, if one sets the detection threshold to be 
$\nuth=4$, the mean number of peaks is 
$N_{\rm peak}(\nunoisy>4)=62$ per 10 deg$^2$.
Among the 62 peaks,  23 are due to real halos with $\nunfw>4$,  
the 13  peaks are halos with $3<\nunfw<4$ and the remaining 26 peaks are 
likely to be either halos with $\nunfw<3$ or false peaks due to the noise.
On the other hand, if one adopts a higher detection threshold
$\nuth=5$,  the mean number of peaks is reduced to
$N_{\rm peak}(\nunoisy>5)=25$ per 10 deg$^2$, 
however the completeness is improved significantly:
in fact it is found that the 19 (3) peaks out of the 25 are signals from 
real halos with $\nunfw>4$ ($3<\nunfw<4$).

Figure \ref{fig:scatter-noisy} shows the scatter plot for the class-(i) 
and class-(ii) halos in the redshift--halo mass plane 
as in Figure \ref{fig:scatter-noisefree}.
{}From comparison with Figure \ref{fig:scatter-noisefree}, 
one may find that more halos pass a certain detection threshold than 
in the noise-free map, because of the systematic bias caused by the noise.
Another important point revealed in this figure is that the peak height in 
the noisy $\kappa$ map is a poor estimator of the halo mass.
For example, masses of the halos detected with $4<\nunoisy<5$ span
over an order of magnitude. 
Even in a narrow redshift range such as $0.3<z<0.4$,
there still exits a significant scatter in the halo mass.
Note that we do not plot halos with $\nunfw<3$, 
which may cause a larger scatter toward small mass ranges. 
It should be also noted that in the distribution of the missing halos
(lower-right panel) no strong tendency with redshift is seen, 
indicating that the completeness would not vary significantly with redshift.
This is largely due to our definition of the completeness,
that is, we define the completeness so that the effect of the selection
function, which very strongly depends on the redshift, is decoupled.

So far we have examined the relation between halos and peaks in a $\kappa$ map
in terms of the peak height instead of the halo mass because the
lensing observable is determined by combination of the halo
mass, the redshift and the noise level, rather than by the halo mass alone. 
Nevertheless, it is interesting to see what mass range of halos produce 
the peaks in the noisy $\kappa$ maps, which is shown in 
Figure \ref{fig:halomass-noisypeak} as in Figure
\ref{fig:halo-noisypeak}. 
Again only halos with $\nunfw>3$ are considered.
As can be seen, the peaks with a given height
are caused by halos with a wide range of masses.
We define the number fraction on the analogy of the completeness 
defined by eq.~(\ref{completeness}) but replacing the 
minimum peak height with the minimum halo mass:
\begin{eqnarray}
\label{fm}
f_m(\nunoisy>\nuth|M_{\rm halo}>M_{\rm min})=
\quad \quad \quad \quad &&\nonumber\\
{{N_{\rm i}(\nunoisy>\nuth|M_{\rm halo}>M_{\rm min})} 
\over {N_{\rm i+ii}(M_{\rm halo}>M_{\rm min})}},&&
\end{eqnarray}
where $N_{\rm i+ii}(M_{\rm halo}>M_{\rm min})$ is the number of 
(class-(i) and (ii)) halos with mass greater than $M_{\rm min}$,
and $N_{\rm i}(\nunoisy>\nuth|M_{\rm halo}>M_{\rm min})$ is 
the number of halos with mass greater than $M_{\rm min}$ having
peak height greater than $\nuth$.
Figure \ref{fig:fm} shows this estimate as a function of $\nuth$.
It is apparent that $f_m$ declines toward higher detection threshold
and for lower minimum mass.
For detection threshold $\nuth=4$, more than 70 percent of halos with 
$M_{\rm halo}>2\times 10^{14}h^{-1}M_\odot$ are identified, while 
about half of halos with
$M_{\rm halo}>5\times 10^{13}h^{-1}M_\odot$ is identified,

\section{Summary and discussion}
\label{sec:summary}

We have investigated various aspects of the weak lensing cluster 
surveys employing both analytic descriptions of dark matter halos 
and the mock data of weak lensing surveys generated from numerical 
simulations. For the latter, we combined weak lensing ray-tracing 
through the mass distribution and the dark matter halo 
catalogs. Our major findings are summarized as follows.

(1) In \S \ref{sec:models}, we examined the expected properties 
of weak lensing halos using the analytic descriptions of the 
dark matter halos, including 
the universal density profile (Navarro et al.~1996; 1997) and the 
Press-Schechter halo mass function (Press \& Schechter 1974: 
Sheth \& Tormen 1999).
We found that the Gaussian smoothing with $\theta_G\simeq 1$ arcmin
gives the largest expected weak lensing halo counts (which may however 
depend on the source redshift and noise properties). We computed the 
selection function of the weak lensing cluster survey and examined, 
in detail, the mass and redshift distribution of the weak lensing 
halos. It was shown that the detectability of halos depends not only
on the mass but also strongly on the redshift.

(2) In \S \ref{halocatalog}, we compared the model prediction of 
the weak lensing halo counts developed by Kruse \& Schneider (2000) 
and Bartelmann et al.~(2001) with the halo counts in the mock 
catalog, and found a good agreement.
We also found a large scatter in the numbers of weak lensing halos
within 4 square degree field, which is larger than the Poisson 
fluctuation. This can be explained by the strong clustering of 
massive halos. 

(3) In \S \ref{sec:result-1}, we tested the model prediction of 
the peak counts against the mock weak lensing data. It was found 
that a systematic bias is induced by the noise due to intrinsic 
galaxy ellipticities. This bias increases the peak counts, and 
hence the model prediction underestimates the counts.
We developed a correction scheme in an empirical manner in 
Appendix \ref{sec:correction}, and showed that the improved 
model reproduces the peak counts reasonably well.

(4) In \S \ref{sec:halo-lens}, using the mock weak lensing data 
combined with the halo catalog, we examined the matching between 
halos and peaks identified in the noise-free $\kappa$ map.
This was done to clarify influence of the individuality of halos 
and the projection effect on a weak lensing halo search. 
We showed that these effects 
cause not only a large scatter in the $\nulens$-$\nunfw$ relation 
but also a systematic bias. The mean and RMS of the differences 
$\nulens-\nunfw$ are $-0.24$ and $1.2$, respectively.
The level of scatter caused by the effects is comparable to that 
due to the noise. We argue that the negative mean value are
due to a population of un-relaxed, less centrally concentrated 
halos and due to the fact that the more than half of the halos 
are elongated to the direction perpendicular to the line-of-sight.
Also the projection of large under-dense regions may partially 
account for the scatter. It was also shown that the chance 
projection of massive halos in the same line-of-sight is very 
rare.

(5) In \S \ref{sec:lens-noisy}, we used the noise-free $\kappa$ 
map and the noisy $\kappa$ map to clarify how the peak distribution
is affected by the noise and how false peaks are generated.
We showed that the $\nulens$-$\nunoisy$ distribution is biased 
toward large $\nunoisy$ values. The mean and RMS of the differences, 
$\nunoisy-\nulens$, among the matched peaks with $\nulens>4$ and 
$\nunoisy>4$ are 0.31 and 0.90, respectively. Thus the noise not 
only generates the scatter but also systematically `boosts' the 
peak heights. We found that almost all the peaks identified in 
the noise-free $\kappa$ map with $\nulens>4$ are identified in the 
noisy $\kappa$ map, thus the noise rarely erases the high peaks.
All the peaks with $\nunoisy>5$ identified in the noisy $\kappa$ 
map are associated with real peaks in the noise-free  $\kappa$ map, 
whereas most of the false peaks (due to the noise) have a 
relatively lower peak height $\nunoisy<4$.

(6) In \S \ref{sec:fof-noisy}, we examined the correspondence between
the halos and peaks identified in the noisy $\kappa$ map. 
In particular, we studied the {\it efficiency} and {\it completeness} 
of the weak lensing halo survey. We found that the detection threshold 
of $\nu_{th}\simeq 4-5$ gives an optimal balance between the efficiency 
and completeness. It was shown that about 81 (62) percent of massive 
halos with $\nunfw>5$ ($>4$) are identified as high peaks with 
$\nuth=4$. This suggests that the completeness of the weak lensing 
cluster survey is reasonably high. Concerning the efficiency, we found 
that for $\nuth=4$ (5), 58\% (82\%) of all the peaks are real signals 
from halos with $\nunfw>3$, while 37\% (63\%) are signals from halos 
with higher expected peak height of $\nunfw>4$. Therefore, it is 
possible to attain relatively high efficiency {\it and} high completeness 
by selecting a moderately high detection threshold.

(7) In \S \ref{sec:fof-noisy}, we conclude that, for detection threshold 
$\nuth=4$, about a half of  halos with 
$M_{\rm halo}>5\times 10^{13}h^{-1}M_\odot$ and 
more than 70 percent of halos with 
$M_{\rm halo}>2\times 10^{14}h^{-1}M_\odot$ indeed produce 
the lensing signals with $\nunoisy\ge 4$.

Weak lensing cluster search technique explored in this paper can be
directly applied to data obtained by on-going/future wide field surveys.
Extending the pilot 2.1 square degree survey (Miyazaki et al.~2002), 
a wide field weak lensing survey (Suprime33, PI: S.~Miyazaki) is being 
carried out with the wide field prime focus camera on Subaru telescope,
Suprime-Cam (Miyazaki et al. in preparation). The Suprime33 will cover 
33 square degrees in total with the limiting magnitude $R=25.5$ (which 
provides $n_g\ga 30$ and $\langle z_s \rangle\simeq 1$). 
For the survey exposure time, 1800 sec, and the field-of-view of the 
Suprime-Cam, 0.25 square degrees, the effective survey cost is 
2.5 hours per 1 square degrees including the overhead. The expected 
number of clusters to be located by the Suprime33 is 70 (120) for the 
threshold $\nuth=5$ ($\nuth=4$). Therefore weak lensing surveys exploiting 
a wide field camera on a large telescope offer a reasonably efficient 
way to locate massive clusters.
The CFHT Legacy Survey\footnote{http://www.cfht.hawaii.edu/Science/CFHLS/}
will significantly enlarge the survey area. Its ``wide'' survey will 
observe 170 square degrees in total with the limiting magnitude 
$i'=25.5$ (which provides $n_g \sim 30$ and $\langle z_s \rangle\simeq 1$).
The expected number of clusters is 360 (600) for $\nuth=5$ ($\nuth=4$),
and thus it will provide invaluable data for studies on large-scale 
structure of the universe. Although we have primarily considered weak 
lensing surveys that are feasible with current ground-based telescopes,
future wide field surveys based on a space telescope will enable to 
accurately measure the shape of distant very small galaxy images, 
which significantly improves the ability of weak lensing cluster 
surveys and allows to detect lower mass and/or higher redshift clusters.

Finally, we note that it is in principle possible to enhance the S/N of 
the detection by choosing a suitable smoothing function
(White et al.~2002; Padmanabhan et al.~2003).
Padmanabhan et al.~(2003) proposed the function of 
$W(\theta)=(1+\theta/\theta_c)^{-2}$ with $\theta_c \sim 1$ arcmin,
which is motivated by the asymptotic behavior of the outer part of 
the projected NFW profile. Such optimized filters may improve the S/N 
if the matter distribution is indeed close to the NFW profile.
On the other hand, its applicability to clusters whose 
mass distribution deviates from the NFW profile remains unclear. 
It needs to be explored how much improvement can be obtained 
both in efficiency and completeness.
Optimizing the smoothing function clearly warrants further studies.

\section*{Acknowledgments}
We are grateful to Bhuvnesh Jain for many valuable discussions.
We would like to thank Satoshi Miyazaki for useful discussions.
We thank the anonymous referee for detailed and constructive
comments on the earlier manuscript which improved various
aspects of the paper.
T.H. and N.Y. thank University of Pennsylvania for the warm hospitality 
during their visit where this work was initiated.  
T.H. and N.Y. acknowledge supports from Japan Society for Promotion of 
Science (JSPS) Research Fellowships. 
The $N$-body simulations used in this work were carried out by the 
Virgo Consortium at the computer center at Max-Planck-Institut, Garching
(http://www.mpa-garching.mpg.de/NumCos). 
Numerical computations presented in this paper were partly
carried out at ADAC (the Astronomical Data Analysis Center) of the
National Astronomical Observatory, Japan.


\appendix

\section{Visual appearances of systems with irregular signals}

\begin{figure}
\begin{center}
\begin{minipage}{8.4cm}
\epsfxsize=8.4cm 
\epsffile{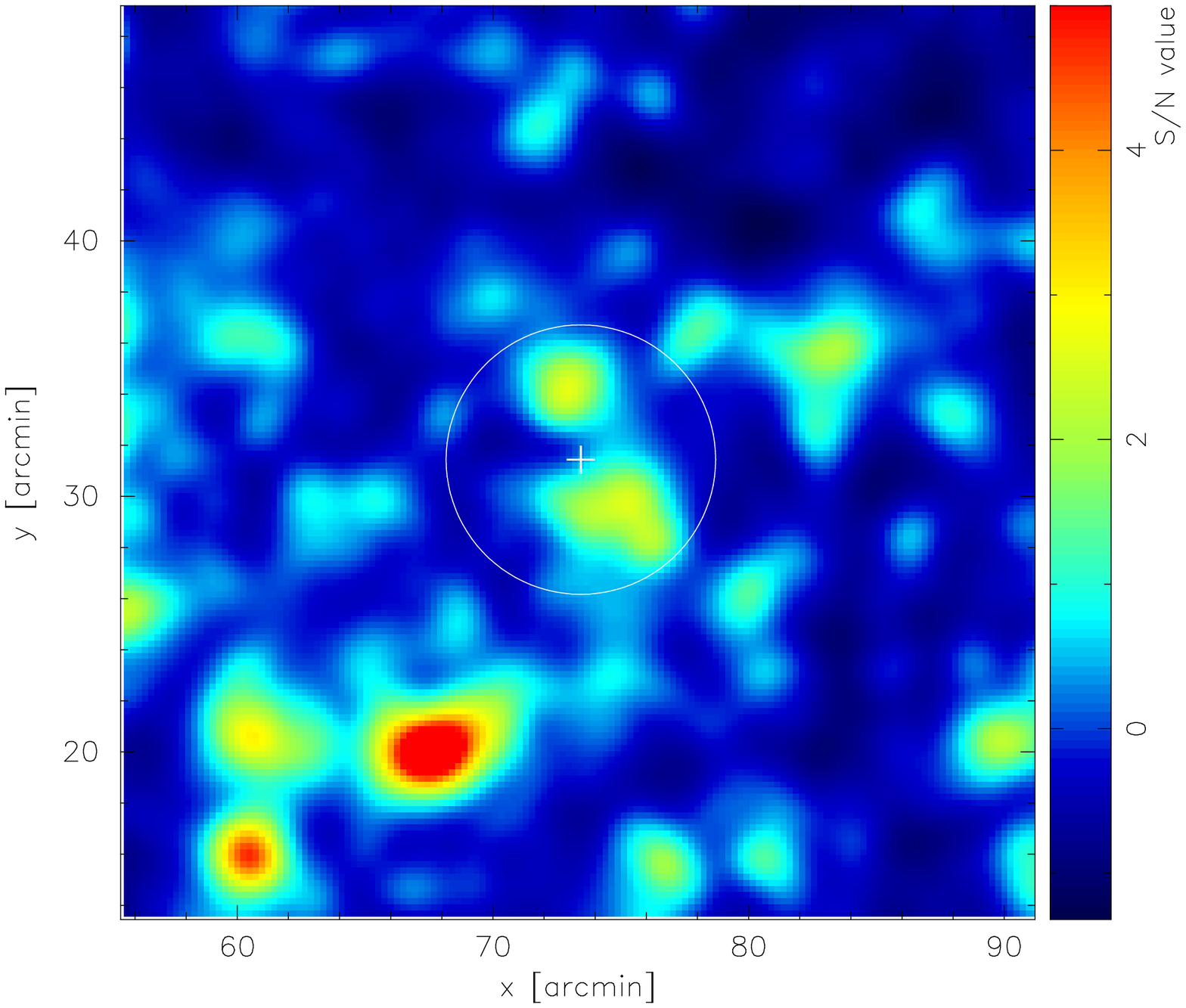}
\end{minipage}
\end{center}
\caption{{\it Missing halo}, an example of a halo with extended matter 
distribution (possibly in the process of merger).
The plus denotes the center-of-mass position of the halo, and the 
circle shows the virial radius.
The map apparently shows very eccentric density distribution.
The halo has $M_{\rm halo}=3.4\times10^{14}h^{-1}M_{\odot}$ at $z=0.37$, 
and the shape parameters are $Q/Q_{\rm NFW}=2.1$, $R=1.4$, and $c/a=0.23$.
The expected peak value is $\nunfw=7.0$, but the $\nu$ value measured in
the lensing map is less than 3.}
\label{fig:missing-halo-1}
\end{figure}

\begin{figure}
\begin{center}
\begin{minipage}{8.4cm}
\epsfxsize=8.4cm 
\epsffile{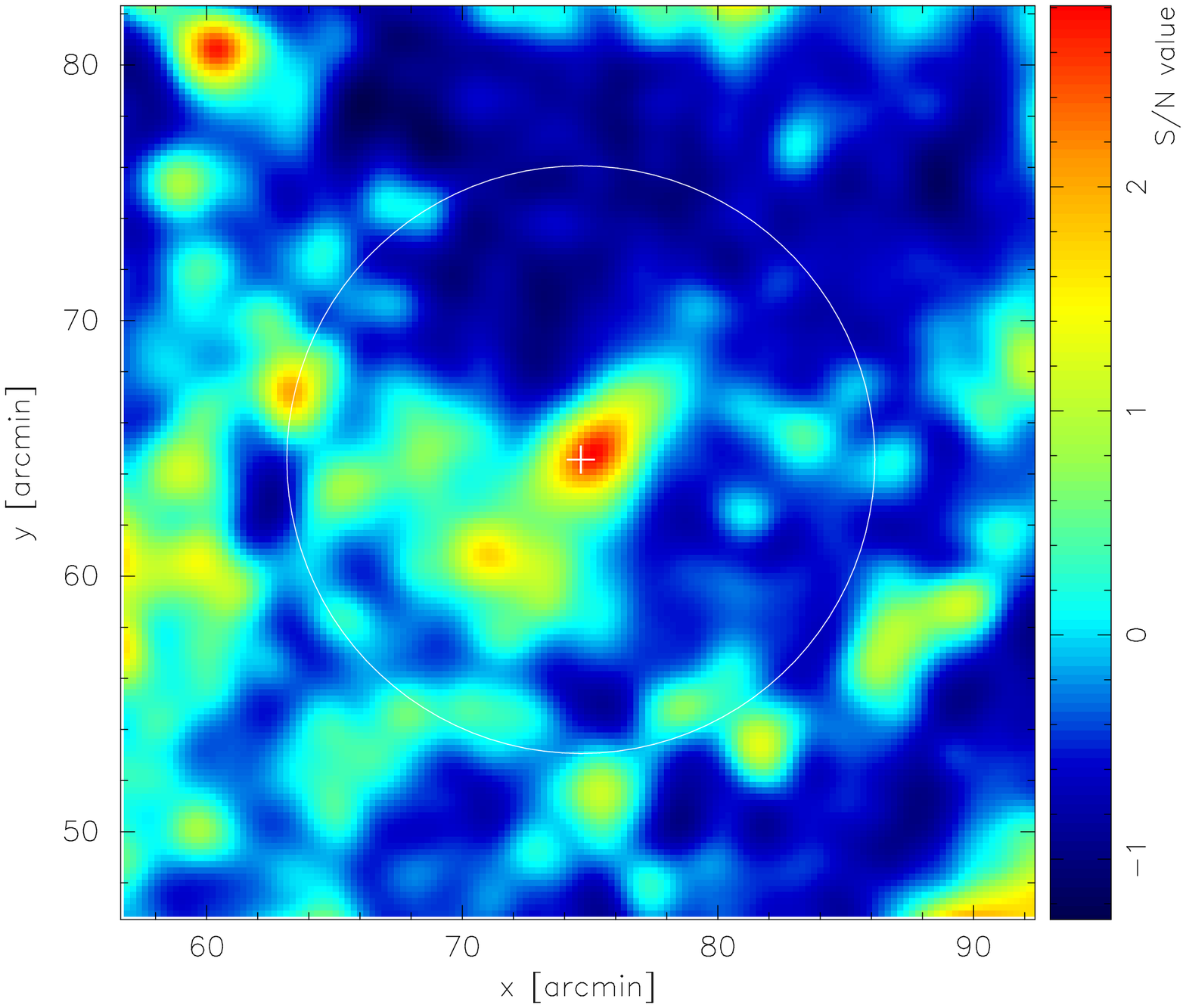}
\end{minipage}
\end{center}
\caption{{\it Missing halo}, an example of low-$z$ un-relaxed system.
The halo ($M_{\rm halo}=2.0\times10^{14}h^{-1}M_{\odot}$, $z=0.13$
and the shape parameters are $Q/Q_{\rm NFW}=1.2$, $R=0.36$, and $c/a=0.41$) 
seems to be not yet reached to a relaxed stage, but probably is a on-going
merger system. The expected peak value is $\nunfw=6.6$ but the actual 
$\nu$ value is less than 3.}
\label{fig:missing-halo-2}
\end{figure}

\begin{figure}
\begin{center}
\begin{minipage}{8.4cm}
\epsfxsize=8.4cm 
\epsffile{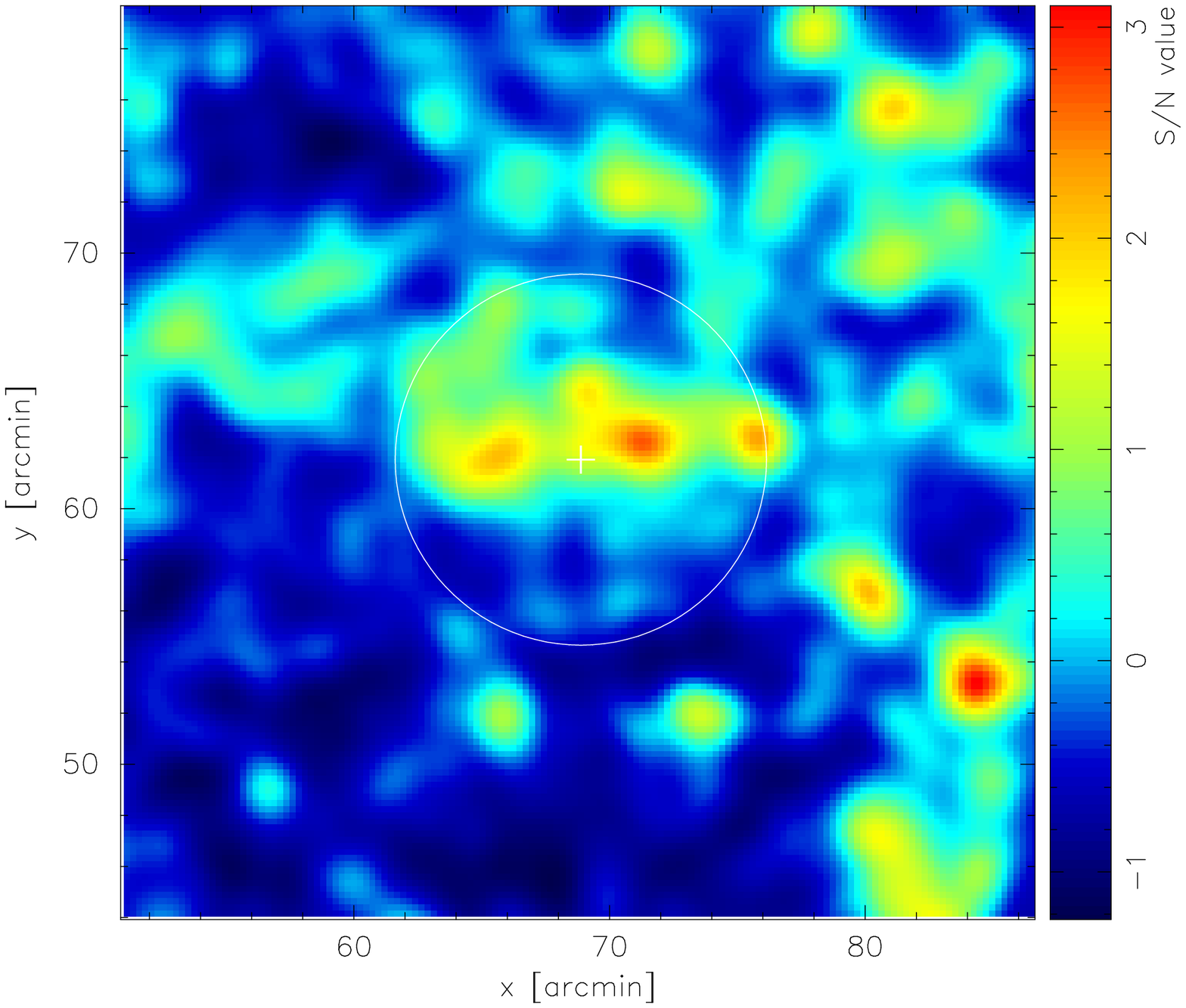}
\end{minipage}
\end{center}
\caption{{\it Missing halo}, an example of an irregular system.
The map apparently shows very eccentric density distribution.
It is very likely that the halo 
($M_{\rm halo}=1.9\times10^{14}h^{-1}M_{\odot}$, $z=0.20$ 
and the shape parameters are $Q/Q_{\rm NFW}=1.3$, $R=0.25$, and $c/a=0.36$.) 
is a on-going merger system.
The expected peak value is $\nunfw=6.1$ but the actual $\nu$ value is 
less than 3.}
\label{fig:missing-halo-3}
\end{figure}

\begin{figure}
\begin{center}
\begin{minipage}{8.4cm}
\epsfxsize=8.4cm 
\epsffile{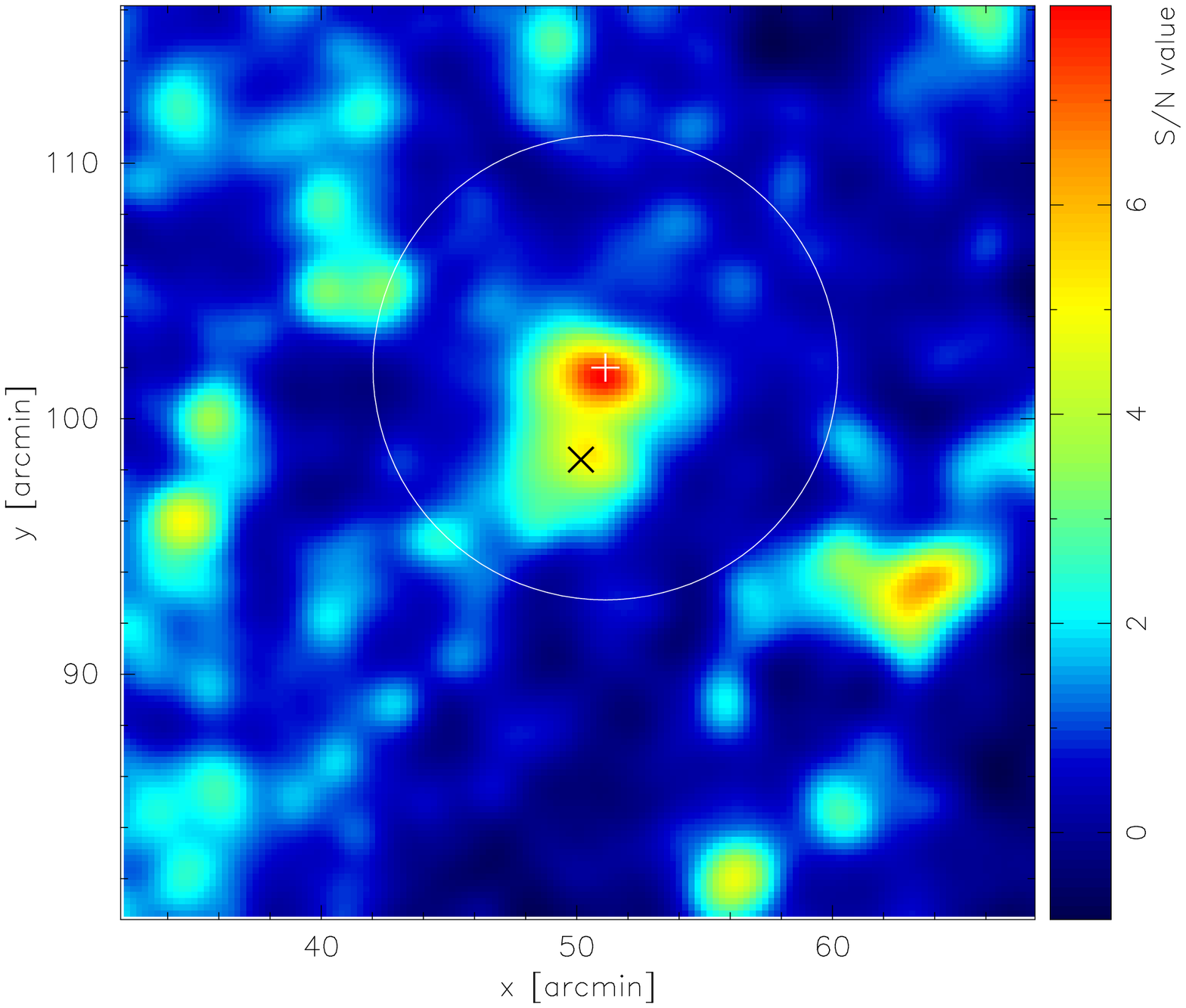}
\end{minipage}
\end{center}
\caption{An example of a {\it local peak} associated 
with the substructure of a massive halo.
The cross denotes the position of the peak with $\nulens=5.3$.
This is the local peak associated with the substructure of the massive 
halo ($M_{\rm halo}=3.5\times10^{14}h^{-1}M_{\odot}$, $z=0.20$, $\nunfw=8.8$
and the shape parameters are $Q/Q_{\rm NFW}=0.94$, $R=0.61$, and $c/a=0.66$).
The plus denotes the center-of-mass position of the halo, and the 
circle shows the virial radius.}
\label{fig:local-peak}
\end{figure}

\begin{figure}
\begin{center}
\begin{minipage}{8.4cm}
\epsfxsize=8.4cm 
\epsffile{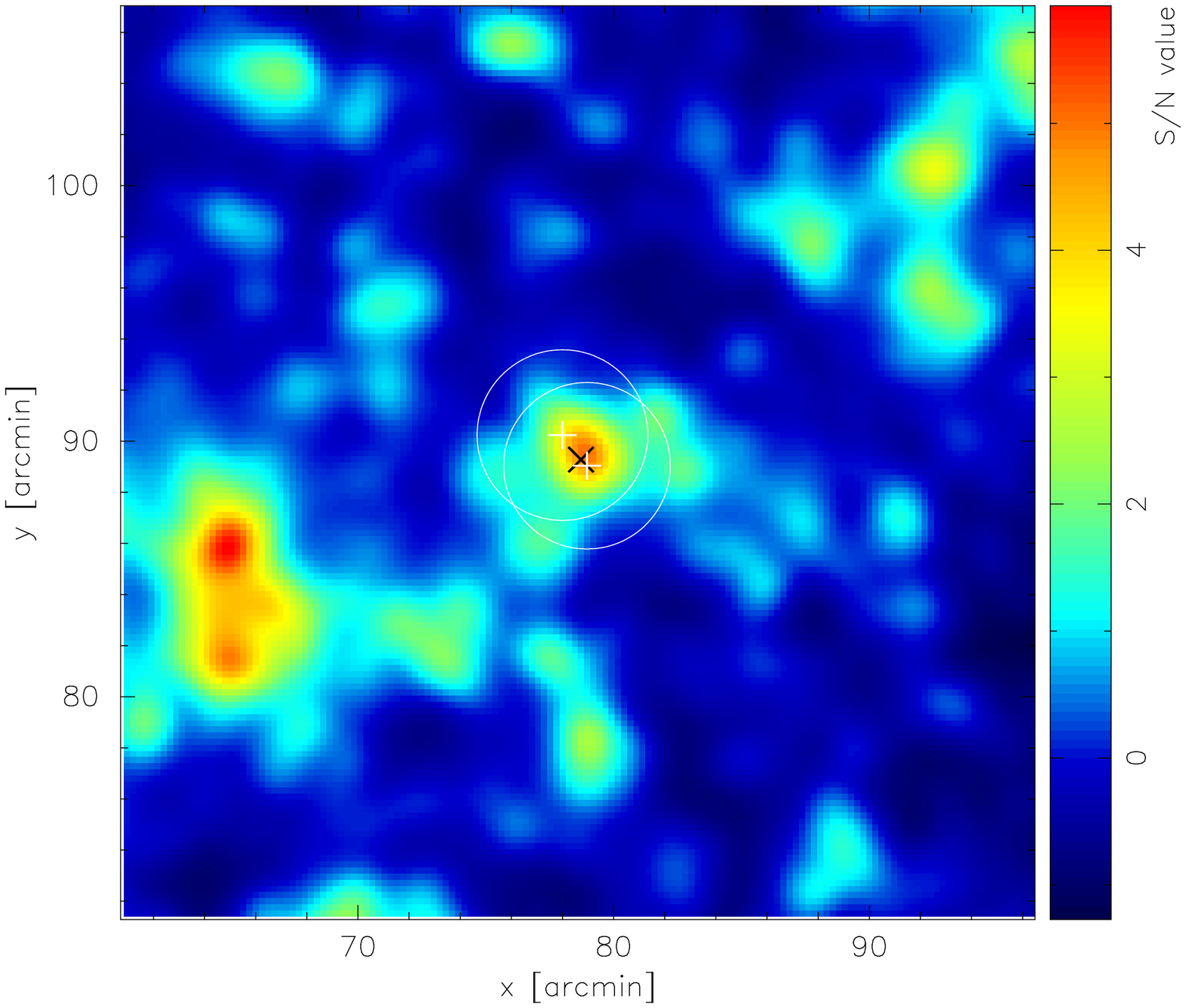}
\end{minipage}
\end{center}
\caption{A very rare example of a {\it chance projection} of 
two moderately massive halos.
The cross denotes the position of the peak with $\nulens=5.1$.
This peak is not generated by a single halo but generated by the 
chance projection of two moderately massive halos:
($M_{\rm halo}=7.5\times10^{13}h^{-1}M_{\odot}$, $z=0.35$) and  
($M_{\rm halo}=5.4\times10^{13}h^{-1}M_{\odot}$, $z=0.68$).
None of two halos alone can generate the lensing signal larger than 
$\nu>3$.}
\label{fig:chance-projection}
\end{figure}

Here we present some $\kappa$ maps, in which {\it missing halos}
(the class-(ii)) or {\it false peaks} (the class-(iii)) with 
high signals ($\nu>5)$ exists, for examples of irregular systems.
Figures \ref{fig:missing-halo-1}-\ref{fig:chance-projection} 
show the (noise-free) convergence maps smoothed with Gaussian window 
function with $\theta_G=1$ arcmin.
Pluses denote the center-of-mass position of halos concerned and the circles
show their virial radius.
Crosses denote the position of peaks.

Figures \ref{fig:missing-halo-1}-\ref{fig:missing-halo-3} show three 
examples of {\it missing halos} with the expected peak value $\nunfw>5$.
Such high $\nunfw$ {\it missing halos} mostly have an un-relaxed appearance 
and are probably on-going merger systems.
Figures \ref{fig:local-peak} and \ref{fig:chance-projection} show
two examples of {\it false peaks} with $\nulens>5$.
Note that {\it false peaks} are rare in noise-free $\kappa$ maps.
Most of {\it false peaks} with $\nulens>5$ are associated with
substructures of massive halos.
Occurrence of the chance projections is very rare, 
we found three cases with $\nulens>5$ in our 30 maps.

\section{Simple correction scheme to the peak counts}
\label{sec:correction}

\begin{figure}
\begin{center}
\begin{minipage}{8.4cm}
\epsfxsize=8.4cm 
\epsffile{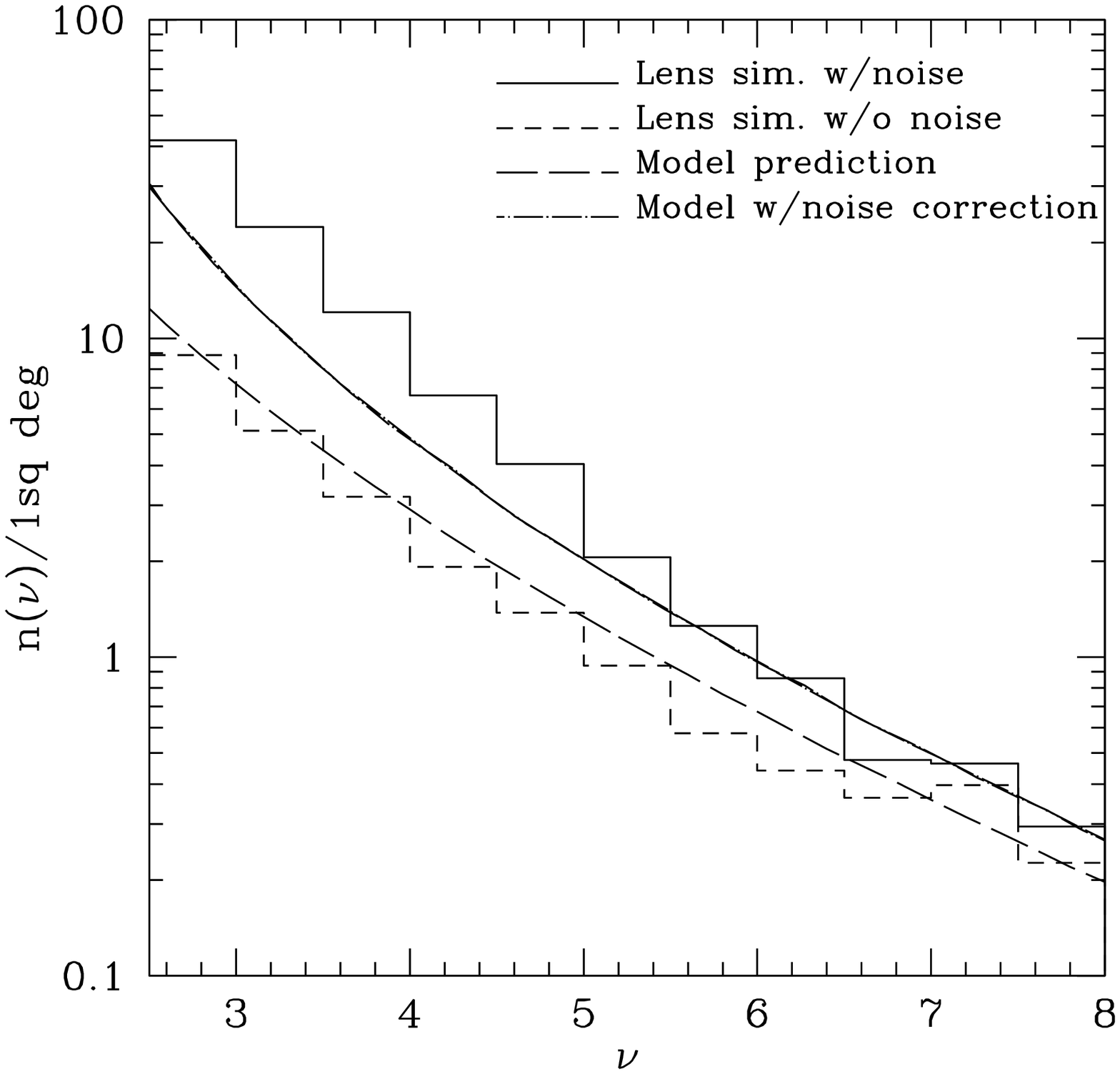}
\end{minipage}
\end{center}
\caption{The number count of the lensing peaks per 1 squire degrees.
The histograms are from the numerical experiments from the noisy 
$\kappa$ maps (solid),
and the noise-free $\kappa$ maps (dashed).
The long-dashed line shows the un-corrected model prediction computed
from eq.~(\ref{defcounts}), while the solid line shows the corrected 
model prediction (see text for detail).}
\label{fig:peakcounts2}
\end{figure}

As shown in \S \ref{sec:result-1} the theoretical model of the peak 
counts given by eq.~(\ref{defcounts}) underestimates the counts obtained 
from the realistic noisy $\kappa$ map by about 20 percent. 
We here discuss sources of this change, and develop the correction 
scheme to the peak counts in an empirical manner.
Here we focus on the high peaks of $\nu>5$.

The sources that modify the peak counts are:
\renewcommand{\labelenumi}{(\alph{enumi})}
\begin{enumerate}
\item The biased relation between $\nunfw$-$\nulens$.
In \S \ref{sec:halo-lens}, it is found that the mean of the differences 
$\nunfw-\nulens$ over class-(i) objects is $-0.24$. 
This means that the peak height computed assuming the NFW profile 
tends to overestimate the real peak height.
The reason for this is discussed in \S \ref{sec:halo-lens}.
\item Spurious peaks generated by the noise. It is found in the 
\S \ref{sec:result-2} that spurious peaks yield about $10\%$ fraction of 
the total number of high peaks ($\nu>5$).
\item The scatter in the relation between $\nunoisy$-$\nulens$ due to 
the noise. As will be shown, this scatter increases the peak counts.
\item The biased relation between $\nunoisy$-$\nulens$.
In \S \ref{sec:lens-noisy}, it is found that the noise produces not only 
the scatter but also the systematic bias, and that the mean of the differences
$\nunoisy-\nulens$ over class-(i) objects is 0.31.
Therefore the peak heights in the noisy $\kappa$ map are statistically 
boosted by the noise. 
The reason for this is discussed in \S \ref{sec:lens-noisy}.
\end{enumerate}

First, we develop the correction scheme for the above point (c). 
To do this, we take a simplified way, employing the assumption
that very high peaks are neither disappeared nor generated 
by the noise but their peak heights are altered by addition of the
noise. In other words, we ignore, for the moment,  spurious peaks due 
to the noise and the other sources. 
Under the above assumption, 
the peak counts in the noisy $\kappa$ map is given by the convolution of
the noise free PDF (\ref{defcounts})
with the Gaussian noise PDF:
\begin{equation}
\label{ncorrect}
n_{\rm noisy}(\nu) = \int d\nu'~ 
{1\over {\sqrt{2 \pi}}}
\exp\left( -{{\nu'^2} \over 2} \right) n_{\rm peak}(\nu-\nu'). 
\end{equation}
We further assume that within a small range of $\nu$ the peak counts in
the original lensing map 
can be approximated by the exponential form
$n_{\rm peak}(\nu)=n_\ast\exp(p\nu)$,
with a constant exponential index of $p$.
As can be seen in Figure \ref{fig:peakcounts2}, this would be a reasonable 
approximation over a range of $3\la \nu\la 8$. 
This approximation allows us to compute 
the integration in eq.~(\ref{ncorrect})
analytically:
\begin{eqnarray}
\label{napprox}
n_{\rm noisy}(\nu) &\simeq& \int d\nu'~ 
{1\over {\sqrt{2 \pi}}}
\exp\left(-{{\nu'^2} \over 2} \right) 
n_\ast\exp\left(p(\nu-\nu')\right)\nonumber\\
&=& \exp\left({{p^2} / 2}\right) n_{\rm peak}(\nu).
\end{eqnarray}
Thus the scatter due to the noise does increase the counts by a factor 
of $\exp(p^2/2)$.

Turn next to the point (a) and (d) in the above list.
{}From the biases in the relations $\nunfw$-$\nulens$
and $\nulens$-$\nunoisy$ found in the mock data, we derive 
an approximate mean relation of $\nunoisy-\nunfw = 0.07$. 
We take this bias by simply shifting the counts
as $n_{\rm noisy}(\nu) \rightarrow n_{\rm noisy}(\nu+0.07)$.

Finally, the point (b) is corrected by simply 
increasing the peak counts by 10 percent.

In Figure \ref{fig:peakcounts2}, the corrected model prediction is 
plotted and is compared with the results from numerical experiments.
To compute the corrected model prediction, $n_{\rm noisy}(\nu)$, we first
computed the uncorrected counts $n_{\rm peak}(\nu)$ at a small interval
of $\delta\nu$ using eq.~(\ref{defcounts}).
Then the local exponential slope of the counts is evaluated by
a finite  differential scheme 
$p(\nu_{i})=[\log n(\nu_{i+1})-\log n(\nu_{i})]/\delta\nu$.
We compute $n_{\rm noisy}(\nu)$ using the eq.~(\ref{napprox}) with 
this $p(\nu_{i})$.
Then, the correction $n_{\rm noisy}(\nu) \rightarrow n_{\rm noisy}(\nu+0.07)$
was made for $p(\nu_{i})$ at every $\nu_{i}$ to correct the 
points (a) and (d).
Finally the counts is increased by 10 percent to correct the point (b).
The corrected prediction agrees with the noisy counts reasonably well 
at least $\nu>5$, where the above approximations are valid.
On lower peak heights $\nu<4$, the prediction is systematically smaller
than the measurement, where false peaks due to the noise have
larger contribution. 
Note that the amplitude of the effects (b) and (d) may depend on the 
noise properties and on the choice of the smoothing scheme.

It is clearly seen in the Figure \ref{fig:peakcounts2} that the 
corrected peak counts has a slope very similar to the un-corrected counts.
Therefore, one may make a simple correction to the model prediction
obtained from eq.~(\ref{defcounts}) by multiplying the factor 
$1.2$, which gives an alternative approximate way to compute the 
corrected peak counts with a reasonable accuracy.

\section{Relations between the halo shape and $\kappa$ amplitude} 
\label{sec:haloshape}

\begin{figure}
\begin{center}
\begin{minipage}{8.4cm}
\epsfxsize=8.4cm 
\epsffile{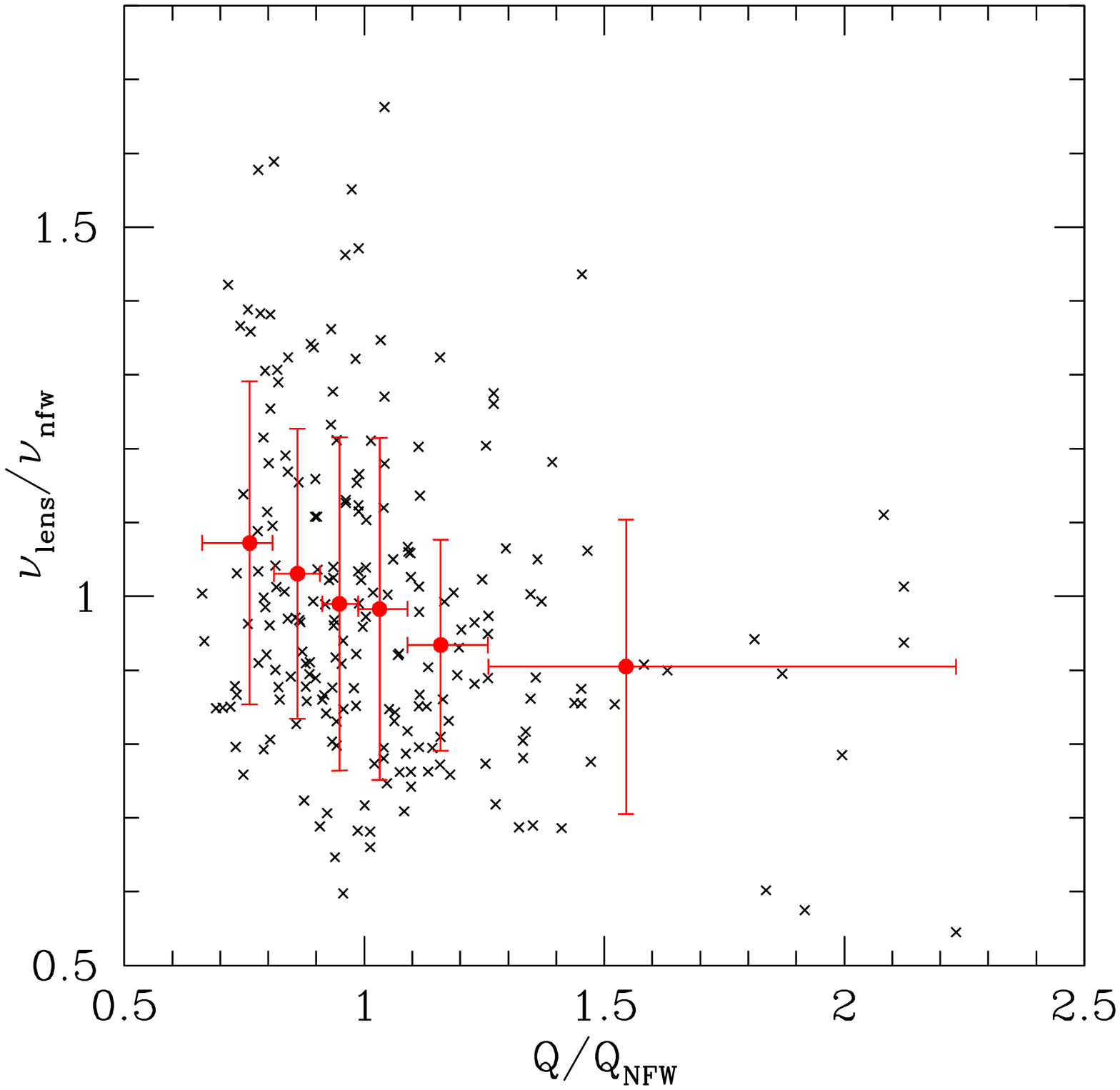}
\end{minipage}
\end{center}
\caption{Shown is the scatter plot of the distribution of the halos 
in $Q/Q_{\rm NFW}$--$\nulens/\nunfw$ plane.
The class-(i) halos having $\nunfw>4$ and $\nulens>4$ only are plotted.
The filled circles show the average over 35 halos within bins whose ranges
are denoted by the $x$-error bars, while
the $y$-error bars denote RMS among the 35 halos.}
\label{fig:q-nu-class1}
\end{figure}

\begin{figure}
\begin{center}
\begin{minipage}{8.4cm}
\epsfxsize=8.4cm 
\epsffile{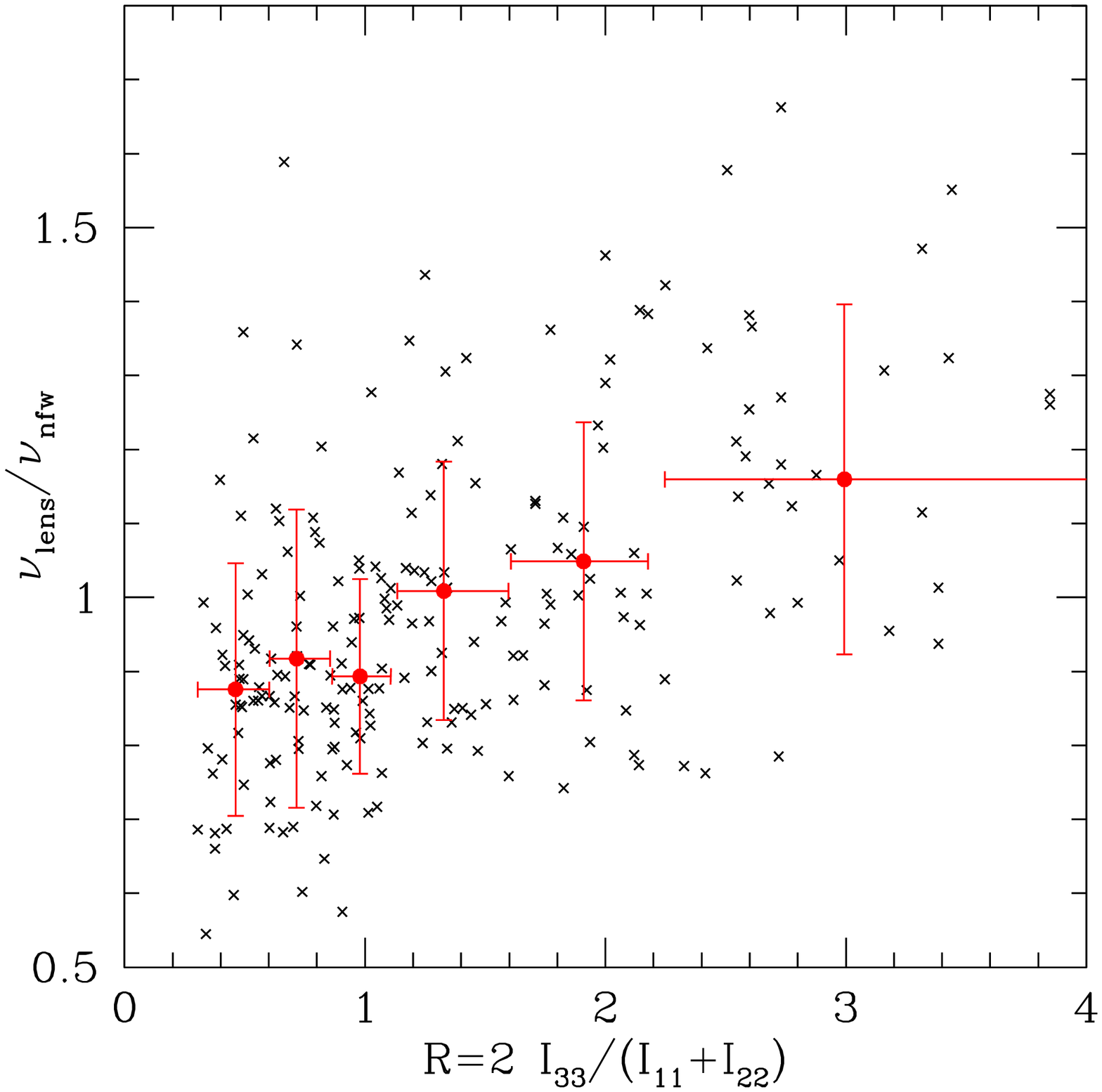}
\end{minipage}
\end{center}
\caption{Same as Figure \ref{fig:q-nu-class1} but for 
$R$--$\nulens/\nunfw$ plane.}
\label{fig:r-nu-class1}
\end{figure}

Here we examine the effect of the individuality in the halo mass
distribution on the amplitude of the $\kappa$ peak.
Figures \ref{fig:q-nu-class1} and \ref{fig:r-nu-class1} show 
how the ratio $\nulens/\nunfw$ varies with 
the halo shape parameters $Q$ and $R$ (see equations 
(\ref{Q}) and (\ref{R}) for the definitions), where we consider 
the class-(i) halos with $\nunfw>4$ and $\nulens>4$ only.
The filled circles show the average over halos sitting in the range
denoted by the $x$-direction bar, 
where each interval is defined so that it contains 35 halos, and the
error bar in $y$-direction denotes the RMS among them.

The correlations between the shape parameters and amplitude of 
the lensing signal are observed in both the figures, though the statistical 
significance is not very high. The anti-correlation in the $\nulens-Q$ is
because more concentrated mass distribution ($Q<1$) compared to the average NFW
profile provides greater lensing signal. The lensing signal is also
amplified if mass distribution within a halo is elongated along the
line-of-sight ($R>1$). However,
we draw attention to the rather large scatters in these relations that 
may arise from the following two sources. First, the shape parameters we
employ characterize rather global mass distribution within a halo and
cannot describe the detailed structures. 
Second, probably more important one is the projection effect due to
different structures along the same line-of-sight of the halo.

\label{lastpage}
\end{document}